\documentclass[final, 5p, times, twocolumn, authoryear]{elsarticle}

\usepackage{cuted}
\usepackage{amssymb}
\usepackage{lipsum}
\usepackage{cuted}
\usepackage{amsmath}
\usepackage{xcolor}  
\usepackage[
    colorlinks=true,
    citecolor=blue,   
    linkcolor=red,    
    urlcolor=magenta 
]{hyperref} 
\usepackage{booktabs}
\newcommand{\cmark}{\checkmark}
\newcommand{\xmark}{--}
\usepackage{hyperref}

\journal{High Energy Astrophysics}

\begin{document}
    \begin{frontmatter}
        \title{VegasAfterglow: A High-Performance Framework for Gamma-Ray Burst Afterglows}
        
        \author[1]{Yihan Wang}
        \fntext[1]{Current affiliation: Department of Astronomy, University of Wisconsin, Madison, WI 53706, USA}
        \ead{wang3697@wisc.edu}
        \author{Connery Chen}
        \ead{connery.chen@unlv.edu}
        \author[2]{Bing Zhang} 
        \fntext[2]{Current affiliation: Department of Physics, University of Hong Kong, Pokfulam Road, Hong Kong, China}
        \ead{bing.zhang@unlv.edu}

        \address{The Nevada Center for Astrophysics, University of Nevada, Las Vegas, NV 89102, USA}
        \address{The Department of Physics and Astronomy, University of Nevada, Las Vegas, NV 89102, USA}

        \begin{abstract}
            Gamma-ray bursts (GRBs) are the most luminous astrophysical transients, known to be associated with core collapse of massive stars or mergers of two compact objects such as two neutron stars. They are followed by multi-wavelength afterglow emission originating from the deceleration of the relativistic jets by the ambient medium. The study of afterglow emission offers crucial insights into the physics of relativistic shocks, the properties of the circumburst environment, the physical and geometrical structure of relativistic jets, as well as the viewing geometry of the observer. We present {\tt VegasAfterglow}, a newly developed, high-performance C++ framework designed for modeling GRB afterglows with flexibility and computational efficiency as key features of design. The framework self-consistently solves forward and reverse shock dynamics and calculates synchrotron (including self-absorption or all spectral regimes) and inverse Compton radiation (including Klein–Nishina corrections); it can handle arbitrary user-defined ambient density profiles, central engine activity histories, viewing angles, and the jet structures of energy, Lorentz factor, and magnetization profiles. It supports both relativistic and non-relativistic regimes and includes lateral jet spreading effects. In this paper, we describe the numerical implementation of the framework and assess its computational performance. Our results demonstrate that {\tt VegasAfterglow} is well-suited for interpreting current and future multi-wavelength observations in the era of multi-messenger astronomy.
        \end{abstract}

    \end{frontmatter}

    \section{Introduction}
    \label{introduction}

    Gamma-ray bursts (GRBs) are brief but intensely luminous flashes of gamma rays from distant galaxies, and remain among the most energetic and enigmatic phenomena in high-energy astrophysics.  Observations revealed two  classes of GRBs with distinct physical origins: those typically with long duration and are associated with the core collapse of massive stars in star-forming regions~\citep{Woosley1993, Galama1998, MacFadyen1999, Woosley2006}, and those typically with short duration and originate from mergers of compact objects such as binary neutron stars (BNS) or neutron star–black hole (NS-BH) systems~\citep{Paczynski1986,  Eichler1989, Paczynski1991, Narayan1992, Gehrels2005, berger_14, ligo_17}. Growing evidence suggests that the duration criterion to separate the two classes becomes blurry and multi-wavelength, multi-messenger observations are essential to reveal the physical origin of a GRB \citep{Zhang2009,Rastinejad2021,Yang2022,Levan2024}.
   
    Regardless of the progenitor and central engine, the emission mechanisms of GRBs are quite generic. In the general framework of GRB emission, the initial hard spike of \textit{prompt} emission is followed by a long-lasting \textit{afterglow} that is detectable from radio to X-rays. The afterglow is widely interpreted as synchrotron radiation emitted by electrons accelerated in shocks driven into the circumburst medium by a relativistic jet~\citep{rees_meszaros_92, meszaros_rees_93, meszaros_rees_97, sari_etal_98, panaitescu_02}. The multi-wavelength properties of the afterglow provide crucial information about the jet structure (e.g., its angular energy, Lorentz factor and magnetization distribution), the ambient density profile, and the microphysical parameters governing shock acceleration and radiation. For comprehensive reviews on GRB physics, see~\citep[e.g.,][]{piran_04, meszaros_06, kumar_15, Zhang2018book}    

    While early GRB modeling often assumed a uniform “top-hat” jet, with constant energy and Lorentz factor within a narrow opening angle and a sharp cutoff beyond~\citep{rhoads_97, sari_etal_99}, this model only accurately describes on-axis observations. At cosmological distances, selection effects favor such detections because the prompt emission is highly beamed. In this scenario, the afterglow light curve (in optical and higher energy bands) typically begins bright and decays monotonically. However, this interpretation breaks down when considering observers located outside the jet core. More realistic models allow for jets with angular structure, where energy and Lorentz factor decrease smoothly with angle from the jet axis~\citep{meszaros_etal_98, dai_gou_01, Rossi2002, zhang_meszaros_02_beaming, kumar_granot_02, Granot2003, Zhang2004Gaussian}. In both scenarios, relativistic beaming makes off-axis afterglows observable, typically exhibiting a delayed rise to peak brightness followed by a decay, though the exact light curve shape and timescale vary depending on the prescribed jet structure.
    
    The detection of GRB 170817A in association with the gravitational wave event GW170817 ushered in the era of multi-messenger astronomy~\citep{ligo_17}.
    For this event, the afterglow exhibited an unusually faint and gradually rising light curve, which provided compelling evidence for an off-axis structured jet~\citep{troja_etal_17, hallinan_etal_17, margutti_etal_17, mooley_etal_18}. The event highlighted the need for modeling frameworks that accurately account for jet structure and observer orientation.

    Over the past two decades, numerous numerical and semi-analytical codes have been developed to model GRB afterglows~\citep[e.g.][]{panaitescu_98,dermer_00,huang_00,zhang_meszaros_01b,granot_sari_02,Granot2003,vaneerten_etal_12,Lei2016,granot_18,ryan_20,wang_etal_24,Nedora2024, Kusafuka2025}. They range from simplified top-hat models to sophisticated hydrodynamical simulations. However, many existing public frameworks lack the flexibility needed to model multi-component or user-defined jet structures, reverse shock emission, synchrotron self-absorption, and inverse Compton (IC) contributions in both the Thomson and Klein-Nishina regimes. In general, analytic and semi-analytic tools are efficient and flexible, allowing for fast light curve generation, but often rely on assumptions like conical geometry, negligible lateral spreading, or approximate blast wave dynamics. Full relativistic hydrodynamic simulations capture complex features such as lateral expansion, shock structure, and jet–medium interaction, but are computationally expensive and impractical for broad parameter inference (e.g., via MCMC sampling). 

    To address these limitations, we present {\tt VegasAfterglow}, a high-performance, modular C++ framework for GRB afterglow modeling that prioritizes both computational efficiency and physical fidelity. Our code self-consistently evolves the dynamics of forward and reverse shocks propagating in a medium with a defined density profile, supports arbitrary user-defined jet structures (including multi-component, asymmetric, and magnetization profiles) with an arbitrary viewing angle, accounts for lateral spreading and energy injection from an arbitrary central engine history, and includes a full treatment of synchrotron (including self-absorption of all spectral regimes) and inverse Compton radiation with optional Klein–Nishina corrections. The framework consistently tracks the evolution of GRB jets into the deep Newtonian regime, enabling accurate modeling across the full afterglow phases.

    In Section~\ref{sec:jets_medium}, we describe the implementation of jets and media with various profiles, including energy injection. In Section~\ref{sec:dynamics}, we discuss relevant equations to characterize the forward and reverse shocks, and discuss the models to account for jet spreading. The subsequent radiation directly depends on the particular shock dynamics of the system, and is discussed in detail in Section~\ref{sec:radiation}. Furthermore, the observed properties of the radiation are highly sensitive to the geometry of the system, as described in Section~\ref{sec:geometry}. In Section~\ref{sec:eg}, we show some examples of afterglow light curves with different setups, benchmark the performance of {\tt VegasAfterglow} and demonstrate its capabilities as a groundbreaking tool for Markov-Chain Monte Carlo simulations. We summarize the key code features and conclude our new findings in Section~\ref{sec:conclusion}.

    \section{Jets and Medium}
    \label{sec:jets_medium}
    {\tt VegasAfterglow} models jet dynamics on a 3D grid of spherical coordinates $(t, \theta, \phi)$, where $t$ is the source frame time, $\theta$ is the polar angle, and $\phi$ is the azimuthal angle. 
    The jet is modeled as a relativistic outflow with a Lorentz factor profile $\Gamma_{0}(\phi, \theta)$, energy profile $\frac{dE}{d\Omega}(\phi, \theta)$, and magnetization profile $\sigma(\phi,\theta)$. In our afterglow calculation, the jet is decomposed into a grid of elements on $\theta$ and $\phi$, and each jet element evolves independently. 

    \subsection{Jet profiles}\label{sec:jet}
    {\tt VegasAfterglow} supports arbitrary user-defined jet profiles for $\Gamma_{0}(\phi, \theta)$, $\frac{dE}{d\Omega}(\phi, \theta)$, and $\sigma(\phi,\theta)$. However, there are a few well-defined axisymmetric jet profiles that are commonly used in the literature. 
    
    \begin{itemize}
        \item The simplest jet profile is the \textbf{top-hat} jet~\citep{rhoads_97, sari_etal_99}, where the Lorentz factor, energy, and magnetization are constant within a certain opening angle $\theta_{\rm c}$ and negligible outside. The top-hat jet is defined as
    \begin{eqnarray}
        \Gamma_{0}( \theta) &=& \left\{
        \begin{aligned}
            &\Gamma_{\rm 0}, \quad \theta\leq\theta_{\rm c} \\
            &1, \quad \theta>\theta_{\rm c}
        \end{aligned}
        \right.\,,\\
        \frac{dE}{d\Omega}(\theta) &=& \left\{
        \begin{aligned}
            &\frac{E_{\rm iso}}{4\pi}, \quad \theta\leq\theta_{\rm c} \\
            &0, \quad \theta>\theta_{\rm c}
        \end{aligned}
        \right.\,,\\
        \sigma(\theta) &=& \left\{
        \begin{aligned}
            &\sigma_{\rm 0}, \quad \theta\leq\theta_{\rm c} \\
            &0, \quad \theta>\theta_{\rm c}
        \end{aligned}
        \right.\,.
    \end{eqnarray}
    \item The \textbf{Gaussian} jet profile~\citep{zhang_meszaros_02_beaming, kumar_granot_02, Zhang2004Gaussian} is defined as
    \begin{eqnarray}
        \Gamma_{0}(\theta) &=& (\Gamma_{\rm 0}(0)-1)\exp\bigg(-\frac{\theta^{2}}{2\theta_{\rm c}^{2}}\bigg)+1\,,\\
        \frac{dE}{d\Omega}(\theta) &=& \frac{E_{\rm iso}(0)}{4\pi}\exp\bigg(-\frac{\theta^{2}}{2\theta_{\rm c}^{2}}\bigg)\,,\\
        \sigma( \theta) &=& \sigma_{\rm 0}\,.
    \end{eqnarray}
    \item The \textbf{power-law} jet profile \citep{meszaros_etal_98, dai_gou_01, rossi_etal_02, zhang_meszaros_02_beaming, Granot2003} is defined as
    \begin{eqnarray}
        \Gamma_{0}(\theta) &=& (\Gamma_{\rm 0}(0)-1)\bigg(1+\bigg(\frac{\theta}{\theta_{\rm c}}\bigg)^k\bigg)^{-1}+1\,,\\
        \frac{dE}{d\Omega}( \theta) &=& \frac{E_{\rm iso}(0)}{4\pi}\bigg(1+\bigg(\frac{\theta}{\theta_{\rm c}}\bigg)^k\bigg)^{-1}\,,\\
        \sigma( \theta) &=& \sigma_{\rm 0}\,.
    \end{eqnarray}
    \end{itemize}

    \subsection{Energy Injection}
    The energy injection mechanism was proposed for modeling the afterglow emission of some GRBs that showed plateau features~\citep{dai_lu_98_injection, zhang_meszaros_01, zhang_06, nousek_06,Lv2015}. Regardless of the specific physical origin, {\tt VegasAfterglow} supports arbitrary user-defined energy injection profiles, $L_{\rm inj}(\phi, \theta, t)$, which can be used to model various injection processes.  
    
    As an example, \texttt{VegasAfterglow} includes several default injection models. One widely used model is the long-lasting Poynting-flux-dominated wind from a spin-down magnetar. This scenario is typically described by a power-law decay function characterized by a timescale $t_{0}$,
    \begin{eqnarray}
        \frac{d L_{\rm inj}}{d\Omega}(\phi, \theta, t) &=& \frac{L_{\rm 0}}{4\pi}\bigg(1+\frac{t}{t_{\rm 0}}\bigg)^{-q}\label{eq:magnetar}\,,\\
        \sigma(\phi, \theta) &=& \infty\,,
    \end{eqnarray}
    where the infinity $\sigma$ profile mimics a Poynting-flux dominated wind, although physically $\sigma$ is a finite but very large value~\citep{Metzger2011, Lei2013}.
    
    \subsection{Medium}\label{sec:medium}
    In principle, the medium can be arbitrarily defined by the user through mass and density functions, $m(\phi,\theta, r)$ and $\rho(\phi, \theta, r)$. However, for most of the GRB afterglow modeling, the medium is assumed to be either homogeneous~\citep{meszaros_rees_97, sari_etal_98} or wind-like~\citep{dai_lu_98_medium, meszaros_rees_99, chevalier_li_00}. The homogeneous medium is defined as
    \begin{eqnarray}
        \rho(\phi, \theta, r) = n_{0} m_{\rm p}   \,, 
    \end{eqnarray}
    where $n_{0}$ is the number density of the interstellar medium. The wind-like medium is defined as
    {\begin{eqnarray}
        \rho(\phi, \theta, r) = \rho_c\bigg(\frac{r}{r_c}\bigg)^{-k}\,,
    \end{eqnarray}
    where $r_{\rm c}$ and $\rho_{\rm c}$ are the characteristic radius and corresponding density of the wind and $k$ is the wind profile index, which is usually adopted as 2 for a constant-velocity wind due to mass flow conservation.}

    \section{Dynamics}\label{sec:dynamics}

    \subsection{General Shock jump conditions}\label{sec:jump}
    The analytical shock jump conditions for an unmagnetized cold upstream have been well established~\citep{blandford_mckee_76, SariPiran1995}.
    For magnetized upstream flows, \citet{kennel_coroniti_84, zhang_kobayashi_05} derived analytical solutions in the highly relativistic regime, where the Lorentz factor between the upstream and downstream is much greater than unity. In contrast, for magnetized non-relativistic shocks, the jump conditions have traditionally been solved numerically. We find that such numerical solutions can become unstable in the regime of low magnetization and small Lorentz factors. In this section, we present a general analytical solution for the shock jump conditions that is valid for an arbitrary Lorentz factor and magnetization parameter.

    The general shock jump conditions are written in the shock frame (denoted by the subscript $s$), with the upstream and downstream regions labeled by subscripts $u$ and $d$, respectively. The shock-frame four-velocities of the upstream and downstream fluids are $u_{\rm us}$ and $u_{\rm ds}$, respectively. In the rest frame of the shock, the continuity equation is given by
        \begin{eqnarray}
            n_{\rm u} u_{\rm us} = n_{\rm d} u_{\rm ds}\,,\label{eq:continuous}
        \end{eqnarray}
    where $n_{\rm u}$ and $n_{\rm d}$ are the proton number densities in the upstream and downstream, respectively. Combined with magnetic flux conservation, energy conservation, and enthalpy conservation
    across the shock front, one can get (see \citealt{zhang_kobayashi_05, Zhang2018book}),
    \begin{eqnarray}
        &&u_{\rm us} = u_{\rm ds}\Gamma_{\rm ud}+\sqrt{(u_{\rm ds}^{2}+1)(\Gamma_{\rm
        ud}^{2}-1)}\label{eq:us}\,,\\
        &&Au_{\rm ds}^6 + Bu_{\rm ds}^4 +Cu_{\rm ds}^2+D
        = 0\,,\label{eq:ds}
    \end{eqnarray}
    where
    \begin{eqnarray}
        A &=& \hat\gamma_{\rm ud}(2-\hat\gamma_{\rm ud})(\Gamma_{\rm ud}-1)+2\,,\\
        B &=& -(\Gamma_{\rm ud}+1)[(2-\hat\gamma_{\rm ud})(\hat\gamma_{\rm ud}\Gamma_{\rm ud}^2+1)+\hat\gamma_{\rm ud}(\hat\gamma_{\rm ud}-1)\Gamma_{\rm ud}]\sigma_{\rm u}
        \nonumber\\ &&-(\Gamma_{\rm ud}-1)[\hat\gamma_{\rm ud}(2-\hat\gamma_{\rm ud})(\Gamma_{\rm ud}^2-2)+(2\Gamma_{\rm ud}+3)]\,,\\
        C &=& (\Gamma_{\rm ud}+1)[\hat\gamma_{\rm ud}(1-\hat\gamma_{\rm ud}/4)(\Gamma_{\rm ud}^2-1)+1]\sigma_{\rm u}^2\nonumber\\
        &&+(\Gamma_{\rm ud}^2-1)[2\hat\gamma_{\rm ud}-(2-\hat\gamma_{\rm ud})(\hat\gamma_{\rm ud}\Gamma_{\rm ud}-1)]\sigma_{\rm u}\nonumber\\
        &&+(\Gamma_{\rm ud}+1)(\Gamma_{\rm ud}-1)^2(\hat\gamma_{\rm ud}-1)^2\,,\\
        D &=& -(\Gamma_{\rm ud}-1)(\Gamma_{\rm ud}+1)^2(2-\hat\gamma_{\rm ud})^2\sigma_{\rm u}^2/4\,,\\
        \hat\gamma_{\rm ud} &=& \frac{4\Gamma_{\rm ud}+1}{3\Gamma_{\rm ud}}\,,
    \end{eqnarray}
    where $\hat\gamma_{\rm ud}$ is the adiabatic index in the downstream. One can prove that the discriminant of the cubic polynomial in Equation~\ref{eq:ds} is always negative, guaranteeing three real solutions for $u_{\rm ds}^2$,
    \begin{eqnarray}
        u_{\rm ds}^2 = 2\sqrt{\frac{-P}{3}}\cos\bigg(\frac{1}{3}\arccos(\frac{3Q}{2P}\sqrt{\frac{-3}{P}})-\frac{2j\pi}{3}\bigg)-\frac{\mathcal{B}}{3}\,,
    \end{eqnarray}
    where $j=0,1,2$ and
    \begin{eqnarray}
        \mathcal{B} &=& \frac{B}{A}, \mathcal{C} = \frac{C}{A}, \mathcal{D} = \frac{D}{A}\,,\\
        P &=& \mathcal{C} - \frac{\mathcal{B}^{2}}{3}\,,\\ 
        Q &=& \frac{2\mathcal{B}^{3}}{27}-\frac{\mathcal{BC}}{3}+\mathcal{D}\,.
    \end{eqnarray}
    Of these, two are unphysical: the $j=0$ solution is excessively large, while the $j=2$ solution is too small, with both violating energy conservation across the shock front. The only physically valid solution corresponds to $j=1$,
    \begin{eqnarray}
        u_{\rm ds}^2 = 2\sqrt{\frac{-P}{3}}\cos\bigg[\frac{1}{3}\arccos\bigg(\frac{3Q}{2P}\sqrt{\frac{-3}{P}}\bigg)-\frac{2\pi}{3}\bigg]-\frac{\mathcal{B}}{3}\,.\label{eq:u_down}
    \end{eqnarray}
    This is the general downstream four-velocity solution for arbitrary Lorentz factors and magnetizations. For an unmagnetized upstream ($\sigma_{\rm u}=0$), the solution is reduced to
    \begin{eqnarray}
        u_{\rm ds}^2 = \frac{(\Gamma_{\rm ud}-1)(\hat\gamma_{\rm ud}-1)^{2}}{\hat\gamma_{\rm
        ud}(2-\hat\gamma_{\rm ud})(\Gamma_{\rm ud}-1)+2}\,, \\
        u_{\rm us}^2 = \frac{(\Gamma_{\rm
        ud}-1)(\hat\gamma_{\rm ud}\Gamma_{\rm ud}+1)^{2}}{\hat\gamma_{\rm ud}(2-\hat\gamma_{\rm
        ud})(\Gamma_{\rm ud}-1)+2}\,,
    \end{eqnarray}
    and Eqn~\ref{eq:continuous} becomes
    \begin{eqnarray}
        n_{\rm d}=\frac{\hat\gamma_{\rm ud}\Gamma_{\rm ud}+1}{\hat\gamma_{\rm ud}-1}n_{\rm u}=4\Gamma_{\rm ud}n_{\rm u}\,.
    \end{eqnarray}

    \begin{figure*}
        \centering
        \includegraphics[width=\textwidth]{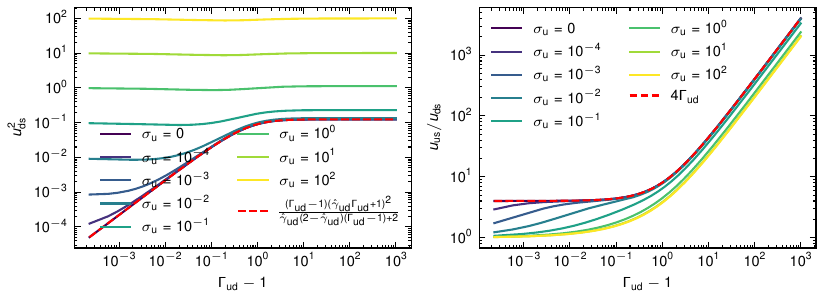}
        \caption{\textit{Left panel:} analytical solution of downstream four velocity $u_{\rm ds}$ as a function of arbitrary relative upstream/downstream Lorentz factor $\Gamma_{\rm ud}$ and upstream magnetization $\sigma_{\rm u}$ (Equation~\ref{eq:u_down}). \textit{Right panel:} upstream-downstream four velocity ratio that gives $n_{\rm d}/n_{\rm u}$.}
        \label{fig:u_down}
    \end{figure*}
    The left panel of Figure~\ref{fig:u_down} shows the analytical solution for the downstream four-velocity, $u_{\rm ds}$, as a function of $\Gamma_{\rm ud}$ for different $\sigma_{\rm u}$, while the right panel shows the upstream to downstream four-velocity ratios.

    Given a known relative Lorentz factor, $\Gamma_{\rm ud}$, and once the downstream proton number density $n_{\rm d}$ is determined, the primary objective of the shock dynamics analysis is to compute the co-moving magnetic field strength $B_{\rm d}$ in the downstream region.
    The $\nabla\cdot \vec B = 0$ gives the conservation of magnetic flux for
    ordered co-moving magnetic field $B_{\rm u,o}$ and $B_{\rm d,o}$ in the upstream or
    downstream, respectively,
    \begin{eqnarray}
        u_{\rm us}  B_{\rm u,o, \perp} = u_{\rm ds}  B_{\rm d,o, \perp}\,.
    \end{eqnarray}
    Then the downstream co-moving magnetic field can be calculated as
    \begin{eqnarray}
         B_{\rm d, \perp} =  B_{\rm d, o, \perp} +  B_{\rm d, w, \perp} =
        \frac{u_{\rm us}}{u_{\rm ds}} B_{\rm u,o,\perp} +
        \sqrt{8\pi\epsilon_{\rm B}e_{\rm d}}\,,\label{eq:B}
    \end{eqnarray}
    where $B_{\rm d, w}$ is the unordered magnetic field generated in the downstream, and $\epsilon_{\rm B}$ is the conventional magnetic energy fraction due to Weibel instability.\footnote{Note that the additional $B_{\rm d, w}$ does not violate $\nabla\cdot \vec B = 0$ if we assume the magnetic field generated due to Weibel instability is generally randomly isotropic, thus $\langle \vec B_{\rm d, w}\rangle\sim0$. See \cite{ma_zhang_22} for a discussion of shock jump conditions with an arbitrary magnetic oblique angle.} One can also define a general magnetic energy fraction, which writes
    \begin{eqnarray}
        \bar{\epsilon}_{\rm B} = \frac{B_{\rm d}^{2}}{8\pi e_{\rm d}} \sim
        \frac{\sigma_{\rm u}}{3(1+\sigma_{\rm u})} + \epsilon_{\rm B}\,.
    \end{eqnarray}
    For the forward shock where $\sigma_{\rm u}=0$, one obtains the conventional $\epsilon_{\rm B}$, and for the reverse shock where the unordered magnetic field is much smaller than the ordered magnetic field and thus $\epsilon_{\rm B}$ is small, one obtains the definition of $\epsilon_{\rm B, r}$ in \cite{zhang_kobayashi_05}.
    \subsection{Blast Wave Evolution}
    For a given relative Lorentz factor, $\Gamma_{\rm ud}$, between the upstream and downstream, the shock jump condition in Section~\ref{sec:jump} gives the downstream proton number density $n_{\rm d}$ and co-moving magnetic field $B_{\rm d}$, which are required for radiation calculations. The primary application of shock dynamics in GRB afterglows is solving the history of $\Gamma_{\rm ud}$. The relative Lorentz factor is given by \footnote{Note that if $\Gamma_{\rm u}\gg1$, $\Gamma_{\rm d}\gg1$, this is reduced to the widely used form $\Gamma_{\rm ud} \sim \frac{1}{2}(\frac{\Gamma_{\rm u}}{\Gamma_{\rm d}}+\frac{\Gamma_{\rm d}}{\Gamma_{\rm u}})$ for reverse shock calculations. However, the reduced form may not be suitable for a relativistic reverse shock in the thick shell regime where $\Gamma_{\rm d}$ can be small in the later stage.}
    \begin{eqnarray}
        \Gamma_{\rm ud} = \Gamma_{\rm u}\Gamma_{\rm d} - \sqrt{(\Gamma_{\rm u}^2-1)(\Gamma_{\rm d}^2-1)}\,,\label{eq:relative_gamma}
    \end{eqnarray}
    where $\Gamma_{\rm u}$ and $\Gamma_{\rm d}$ are the Lorentz factors of the upstream and downstream, respectively. The upstream Lorentz factor, $\Gamma_{\rm u}$ is given by the jet Lorentz factor, $\Gamma_{\rm j}$, or the medium Lorentz factor, $\Gamma_{\rm m}$, which is already known. So one may need to solve the downstream Lorentz factor by blast wave equations. Generally, as the jet collides with the medium, a pair of shocks (i.e. a the forward shock and the reverse shock) could be generated as long as the reverse shock generation condition is satisfied. Otherwise, only a forward shock may be generated. The upstream/downstream of the forward and reverse shocks divide the jet and medium into four regions: the unshocked medium, upstream of the forward shock (region 1), the shocked medium, downstream of the forward shock (region 2), the shocked jet, downstream of the reverse shock (region 3) and the unshocked jet, upstream of the reverse shock (region 4). The reverse shock will propagate toward the central engine until all the jet material (region 4) is crossed. Meanwhile, the forward shock will continue to propagate through the external medium.

    \subsubsection{Reverse Shock Crossing via pressure balance}
    During the reverse shock crossing phase, one must combine the shock jump conditions of the forward and reverse shocks to obtain the target downstream Lorentz factors, $\Gamma_2$ and $\Gamma_3$. For the forward shock, the upstream is region 1 and the downstream is region 2, thus, using Equation~\ref{eq:continuous},
    \begin{eqnarray}
        n_2 &=& 4\Gamma_{12}n_1\label{eq:begin}\,,\\
        e_2 &=& (\Gamma_{12}-1)n_2m_{\rm p}c^2\,,
    \end{eqnarray}
    where $n_{1}= n_{\rm medium}$ is provided in Section~\ref{sec:medium},
    $\sigma_{1}=0$, and $\Gamma_{1}=1$.

    For the reverse shock, the upstream is region 4 and the downstream is region
    3, thus,
    \begin{eqnarray}
        n_3 &=& \frac{u_{4s}}{u_{3s}}n_4 = \bigg(\Gamma_{43}+\frac{\sqrt{(u_{\rm
        3s}^{2}+1)(\Gamma_{\rm 43}^{2}-1)}}{u_{3s}}\bigg)n_4\,,\\ 
        e_3 &=& (\Gamma_{43}-1)n_3m_{\rm p}c^2\,,\\
        {B_{3}^{2}} &=& \frac{u_{4s}^{2}}{u_{3s}^{2}}{B_{4}^{2}}\,,\label{eq:end}
    \end{eqnarray}
    where
    \begin{eqnarray}
        n_4 &=& \frac{E_{\rm jet}}{4\pi r^2\Delta^\prime\Gamma_{4}m_{p}c^{2}(1+\sigma_{4})}\,,\\
        {B_{4}^{2}} &=&4\pi\sigma_4 n_4 m_{\rm p}c^2\,,
    \end{eqnarray}
    where $\Delta^\prime$ is the shell width in the co-moving frame. The force balance at the discontinuity gives \citep{SariPiran1995,zhang_kobayashi_05},
    \begin{eqnarray}
        p_2 &=& (\hat\gamma_{12}-1)e_2=(\hat\gamma_{43}-1)e_3+
        \frac{B_{3}^{2}}{8\pi} = p_3\,.\label{eq:fr-balance}
    \end{eqnarray}
    Combine Equation~\ref{eq:begin} and \ref{eq:end} with this discontinuity condition to get
    \begin{eqnarray}
        1 = \frac{p_2}{p_3} \sim \frac{n_1(\Gamma_{12}^2-1)}{n_4(1+\sigma_4)(\Gamma_{43}^2-1)}\,.\label{eq:fr-eqn}
    \end{eqnarray}
    Since $\Gamma_1$, $\Gamma_4$, $n_1$ and $n_4$ are known, using Equation~\ref{eq:relative_gamma} and the fact that $\Gamma_2=\Gamma_3$, one eventually obtains $\Gamma_2$ ($\Gamma_3$) during the reverse shock crossing phase. 

    If the pressure in region 4 (the unshocked jet) is too strong, the reverse shock cannot be formed. Using Equation~\ref{eq:fr-balance}, one gets the reverse shock generation condition $e_{3}>0$, which is
    \begin{eqnarray}
        \sigma_4&\lesssim&8(\hat\gamma_{12}-1)\Gamma_{12}(\Gamma_{12}-1)\frac{n_{1}}{n_{4}}=\frac{8}{3}(\Gamma_{12}^2-1)\frac{n_{1}}{n_{4}}\nonumber\\
        &\lesssim&\frac{8}{3}(\Gamma_{3}^2-1)\frac{n_{1}}{n_{4}}\sim \frac{8}{3}(\Gamma_{4}^2-1)\frac{n_{1}}{n_{4}}\label{eq:rs-gen}\,,
    \end{eqnarray}
    given the condition that $\epsilon_{\rm B, r}\ll 1$, $u_{3s}\sim u_{4s}$,
    and $\Gamma_{12}=\Gamma_{2}=\Gamma_{3}\sim\Gamma_{4}$. 

    \subsubsection{Reverse Shock Crossing via energy conservation}
    
    Equation~\ref{eq:fr-balance} assumes that the Lorentz factor and pressure in regions 2 and 3 are constant (thus the Lorentz factor and pressure immediately behind the shock front in the downstream equal the pressure at the discontinuity). This approximation is valid when the reverse shock crossing time is short and the resulting blast wave is thin. However, for a long-duration reverse shock that produces a thick blast wave, this assumption is invalid and the pressure-balance method can lead to significant violations of energy conservation.

    \begin{figure}
        \centering
        \includegraphics[width=.45\textwidth]{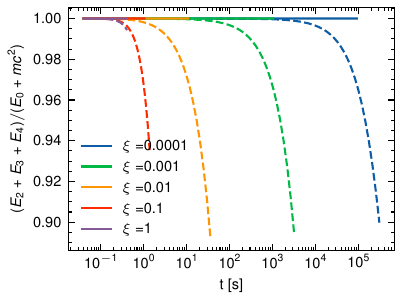}
        \caption{Energy conservation check during the reverse shock crossing phase for the classical forward/reverse shock model (dashed lines) that assumes constant bulk Lorentz factor and pressure in the blast wave, and for our effective mechanical model introduced in this work (solid lines). $\xi$ is the initial shell thickness parameter for the jet.}
        \label{fig:rs-correction}
    \end{figure}
    The dashed lines in Figure~\ref{fig:rs-correction} show the energy conservation during the reverse shock crossing based on the above treatment for different shell thickness regimes. It is indicated that for a thick shell ($\xi\ll1$) this treatment leads to a $\sim10\%$ energy conservation issue, which will become even worse for a thicker shell.

    To solve this problem, a mechanical model \citep{beloborodov_uhm_06,uhm_beloborodov_07,uhm_11} has been developed to ensure energy conservation during the reverse shock crossing phase, where a pressure profile in the blast wave has been assumed, so Equation~\ref{eq:fr-eqn} is no longer valid. \cite{beloborodov_uhm_06} proved that $p_2/p_3$ can be as large as $3$ for a stellar wind medium and that the magnetization of the jet further complicates the mechanical model \citep{Ai2021}.

     Rather than adopting existing mechanical models, we employ an energy conservation method to solve the shock dynamics during the reverse shock crossing phase. { During the shock crossing phase, the total energy in the blast wave, including region 2 and region 3, can be expressed as
\begin{eqnarray}
    \Gamma_2m_2c^2+\Gamma_{\rm eff,2}U_2+\Gamma_3m_3c^2+\Gamma_{\rm eff,2}U_2\label{eq:e_blast}
\end{eqnarray}
where
\begin{eqnarray}
    \Gamma_{\rm eff,2}&=&\frac{\hat{\gamma}_{2}\Gamma_{2}^{2}-\hat{\gamma}_{2}+1}{\Gamma_{2}}\,,\\
    \Gamma_{\rm eff,3}&=&\frac{\hat{\gamma}_{34}\Gamma_{3}^{2}-\hat{\gamma}_{34}+1}{\Gamma_{3}}\,.
\end{eqnarray}

During the shock crossing phase, the forward shock propagates into the interstellar medium and compresses the material into region 2, while the reverse shock compresses the material from region 4 into region 3. Therefore, the blast wave dynamics equation can be written as
\begin{eqnarray}
    d[\Gamma_2m_2c^2+\Gamma_{\rm eff,2}U_2+\Gamma_3m_3c^2+\Gamma_{\rm eff,2}U_2]=c^2dm_2\nonumber\\
    +c^2\Gamma_4dm_3 +\Gamma_{\rm eff,2}dU_{\rm rad,2} +\Gamma_{\rm eff,3}dU_{\rm rad,3} + L_{\rm inj}dt
\end{eqnarray}
where $dU_{\rm rad,2}$ and $dU_{\rm rad,3}$ are the internal energy losses due to radiation in regions 2 and 3, respectively, and $L_{\rm inj}$ is the external energy injection luminosity.  The internal energy evolution is
\begin{eqnarray}
    dU_{\rm i} = dU_{\rm sh,i} + dU_{\rm ad,i} + dU_{\rm rad,i}
\end{eqnarray}
which includes shock heating $dU_{\rm sh,i}$, adiabatic cooling $dU_{\rm ad,i}$, and radiation $dU_{\rm rad,i}$, where
\begin{eqnarray}
    dU_{\rm sh,2} &=& (\Gamma_2-1)c^2dm_2\\
    dU_{\rm rad,2} &=&-\epsilon_2dU_{\rm sh,2}= -\epsilon_{\rm rad,2}\epsilon_{e,2}dU_{\rm sh,2}\\
    dU_{\rm ad,2} &=& -pdV_2^\prime=-(\hat\gamma_2-1)\frac{U_2}{V_2^\prime}dV_2^\prime\nonumber\\
    &=&-(\hat\gamma_2-1)U_2\bigg(2\frac{dr}{r}+\frac{d\Delta_2^\prime}{\Delta_2^\prime}\bigg)\\
    dU_{\rm sh,3} &=& (\Gamma_{34}-1)c^2dm_3\\
    dU_{\rm rad,3} &=&  -\epsilon_3dU_{\rm sh,3}=-\epsilon_{\rm rad,3}\epsilon_{e,3}dU_{\rm sh,3}\\
    dU_{\rm ad,3} &=& -pdV_3^\prime =-(\hat\gamma_{34}-1)\frac{U_3}{V_3^\prime}dV_3^\prime\nonumber\\
    &=&-(\hat\gamma_{34}-1)U_3\bigg(2\frac{dr}{r}+\frac{d\Delta_3^\prime}{\Delta_3^\prime}\bigg)
\end{eqnarray}
where $V_i^\prime=4\pi r^2\Delta_i^\prime$ is the comoving volume in region $i$\footnote{For adiabatic cooling, we adopt a more accurate description of shell evolution instead of the commonly used $r/\Gamma$ approximation in the literature.}, $\Delta^\prime_i$ is the comoving shell thickness in region $i$, and the radiation efficiency in region $i$ can be expressed as
\begin{eqnarray}
    \epsilon_{\rm rad,i}=\left\{
    \begin{aligned}
        &1,\quad &\gamma_{c,i}<\gamma_{m,i}&\\
        &\bigg(\frac{\gamma_{c,i}}{\gamma_{m,i}}\bigg)^{2-p},\quad &\gamma_{c,i}>\gamma_{m,i}&,\quad p_i>2&
    \end{aligned}
    \right.\,,
\end{eqnarray}
(see the definitions of $\gamma_m$, $\gamma_c$ and $\epsilon_e$ in Section~\ref{sec:radiation}).

To close these equations, we also need the shell evolution equation:
\begin{eqnarray}
    \frac{d\Delta_2^\prime}{dt} &=& c_{s,2}\frac{dt^\prime}{dt}\\
    \frac{d\Delta_3^\prime}{dt} &=& \Gamma_3\frac{\beta_4-\beta_3}{\beta_3}\bigg(\frac{\Gamma_{3}n_{3}}{\Gamma_{4}n_{4}}-1\bigg)^{-1}\frac{dr}{dt}\,,
\end{eqnarray}
where
\begin{eqnarray}
    c_{s,2}=c\sqrt{\frac{\hat\gamma_{2}p}{\rho_0c^2+\frac{\hat\gamma_{2}}{\hat\gamma_{2}-1}p}}=c\sqrt{\frac{\hat\gamma_{2}(\hat\gamma_{2}-1)(\Gamma_{2}-1)}{1+\hat\gamma_{2}(\Gamma_{2}-1)}}
\end{eqnarray}
is the sound speed in region 2, and the mass accumulation equations are
\begin{eqnarray}
    \frac{dm_2}{dt}&=&4\pi r^2\rho_1\frac{dr}{dt}\\
    \frac{dm_3}{dt}&=&4\pi r^2n_3m_p\frac{d\Delta_3^\prime}{dt}\label{eq:r-cross1}
\end{eqnarray}

By assuming $\Gamma_2=\Gamma_3$, ${dr_2}/{dt}={dr_3}/{dt}={c\beta_2}/(1-\beta_2)$ during the shock crossing phase, and putting everything together, one obtains

\begin{eqnarray}
    \frac{d\Gamma_2}{dt}&=&-\frac{(\Gamma_2-1)c^2\frac{dm_2}{dt}+\Gamma_{\rm eff,2}(\frac{dU_{\rm sh,2}}{dt}+\frac{dU_{\rm ad,2}}{dt})}{(m_2+m_3)c^2+\frac{d\Gamma_{\rm eff,2}}{d\Gamma_2}U_2+\frac{d\Gamma_{\rm eff,3}}{d\Gamma_2}U_3}\nonumber\\
    &-&\frac{(\Gamma_2-\Gamma_4)c^2\frac{dm_3}{dt}+\Gamma_{\rm eff,3}(\frac{dU_{\rm sh,3}}{dt}+\frac{dU_{\rm ad,3}}{dt})}{(m_2+m_3)c^2+\frac{d\Gamma_{\rm eff,2}}{d\Gamma_2}U_2+\frac{d\Gamma_{\rm eff,3}}{d\Gamma_2}U_3}\nonumber\\
    &+&\frac{L_{\rm inj}}{(m_2+m_3)c^2+\frac{d\Gamma_{\rm eff,2}}{d\Gamma_2}U_2+\frac{d\Gamma_{\rm eff,3}}{d\Gamma_2}U_3}\label{eq:mech-eqn}\\
    \frac{dU_2}{dt}&=&(1-\epsilon_2)(\Gamma_2-1)c^2\frac{dm_2}{dt}\nonumber\\
    &-&(\hat\gamma_2-1)U_2\bigg(\frac{2}{r}\frac{dr}{dt}+\frac{1}{\Delta_2^\prime}\frac{d\Delta_2^\prime}{dt}\bigg)\\
    \frac{dU_3}{dt}&=&(1-\epsilon_3)(\Gamma_{34}-1)c^2\frac{dm_3}{dt}\nonumber\\
    &-&(\hat\gamma_{34}-1)U_3\bigg(\frac{2}{r}\frac{dr}{dt}+\frac{1}{\Delta_3^\prime}\frac{d\Delta_3^\prime}{dt}\bigg)\label{eq:u3}
\end{eqnarray}

}

Equation~\ref{eq:mech-eqn} describes the blast wave dynamics until the reverse shock has fully propagated through the ejecta, which occurs when
    \begin{eqnarray}
        m_{3}(r_\times) = m_{ 4}
        \equiv \frac{E_{\rm jet}}{\Gamma_{4}c^{2}(1+\sigma_{4})}\,,\label{eq:r-cross2}
    \end{eqnarray}
    where $r_{\times}$ is the reverse shock crossing radius. 

    The solid lines in Figure~\ref{fig:rs-correction} demonstrate that the new treatment conserves energy significantly better during long-lasting reverse shocks with larger shell thicknesses, parameterized as \citep{SariPiran1995}
    \begin{eqnarray}
        \xi=\bigg(\frac{l}{\Delta}\bigg)^{1/2}\Gamma_4^{-4/3}\,,
    \end{eqnarray}
    where $l$ is the Sedov length and $\Delta = cT$, with $T$ being the duration of the burst. This approach effectively functions as a simplified mechanical model: rather than explicitly integrating the pressure, enthalpy, and density profiles within the blast wave, it relies on the total energy in region 2 and region 3. This formulation greatly simplifies the traditional mechanical model while retaining its essential physics.

  \begin{figure*}
        \centering
        \includegraphics[width=\textwidth]{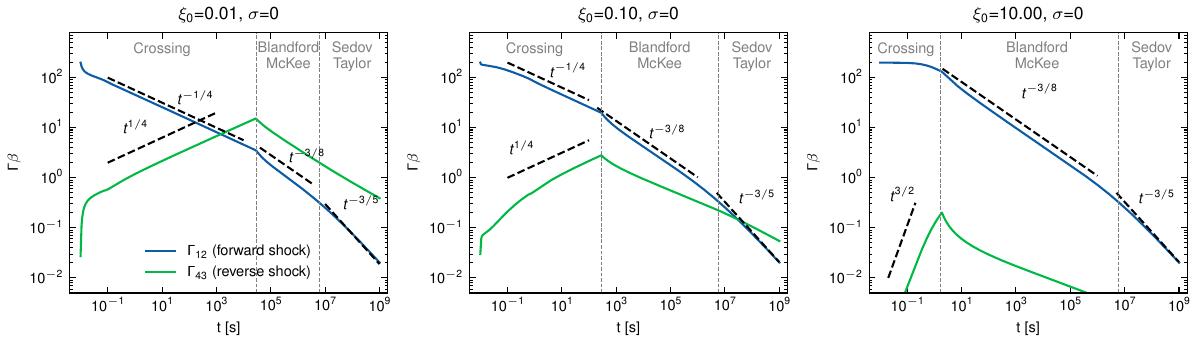}\\
        \includegraphics[width=\textwidth]{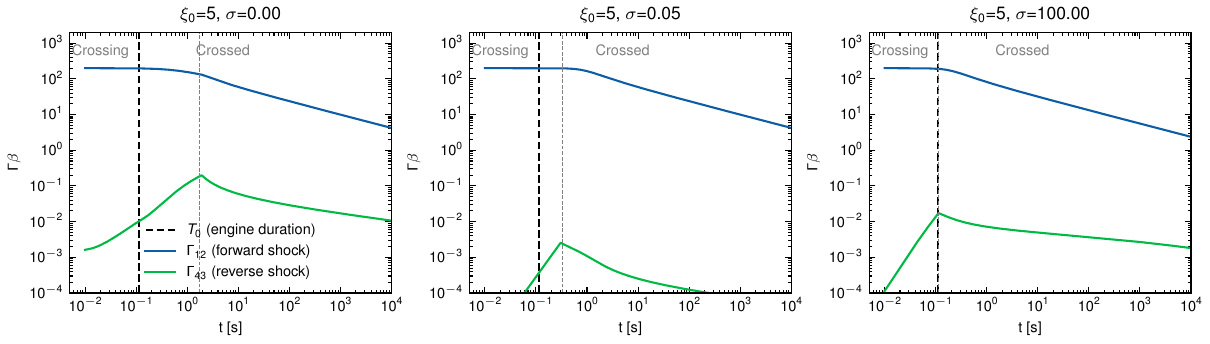}
        \caption{Forward and reverse shock dynamics for varying jet shell thickness ($\xi$) and magnetization ($\sigma$). The upper panels display dynamics for unmagnetized shells, transitioning from the thick-shell regime ($\xi\ll1$) to the thin-shell regime ($\xi\gg1$). The bottom panels show thin-shell cases with different levels of magnetization ($\sigma$).}
        \label{fig:r_dynamics}
    \end{figure*}
    
Figure~\ref{fig:r_dynamics} shows examples of the forward and reverse shock {four-velocity evolution} for different shell thicknesses and magnetizations. The upper panels display the evolution of the four-velocities of the forward shock ($\Gamma_{12}$), reverse shock ($\Gamma_{43}$), and blast wave tail ($\Gamma_3$) for the unmagnetized case in both the thick-shell regime ($\xi_0 \ll 1$) and thin-shell regime ($\xi_0 \gg 1$). 

We remark that in the thin shell regime, where $\xi_0 \gg 1$, the reverse shock remains non-relativistic throughout the time it crosses the entire shell, with a peak Lorentz factor {$\Gamma_{43} \sim 1.02$ ($\Gamma_{34}\beta_{34}\sim0.2$).} This is in contrast to the mildly relativistic value $\Gamma_{43} \sim 7/4$ derived in \cite{SariPiran1995} and \cite{Kobayashi2000}. Several factors contribute to this discrepancy: 

\begin{itemize}
    \item They adopted an inaccurate shock jump condition, $n_3/n_4 = 4\Gamma_{43} + 3$, by assuming $\hat{\gamma}_{34} = 4/3$, which overestimates the shell compression in region 3 (a factor of 7 instead of the correct factor of 4); 

    \item  They neglected the $\hat{\gamma} - 1$ factor in pressure calculations for regions 2 and 3, where $\hat{\gamma}_2 \neq \hat{\gamma}_{43}$; 

    \item  They omitted the term $\left(1 - \frac{\Gamma_{4} n_{4}}{\Gamma_{3} n_{3}}\right)$ in deriving the shock crossing radius, which is approximately $3/4$ in the thin shell regime, as noted in \cite{zhang_kobayashi_05}; 

    \item  As the reverse shock (characterized by $\Gamma_{43}$) strengthens, the bulk Lorentz factor $\Gamma_3$ decreases toward $\sim\Gamma_4/2$ during the crossing. This evolution gradually changes the scaling of $\Gamma_{43} - 1 \propto t^3$ (typical for the thin shell regime) to $\Gamma_{43} - 1 \propto t^{1/4}$ (typical for the thick shell regime), thereby slowing the growth of $\Gamma_{43}$ in the later stages of the crossing (see upper right panel of Figure~\ref{fig:r_dynamics}).
{
    \item The energy-conservation method adopted in this work yields a lower $\Gamma_{43}$ than the traditional pressure-balance method. This difference arises because the commonly used internal energy approximation, $U_3 = (\Gamma_{34}-1)m_3c^2$, overestimates the actual internal energy in region 3. The exact internal energy is given by the integral form of Equation~\ref{eq:u3}, in which the early shock heating is weaker. Moreover, adiabatic cooling further reduces the internal energy relative to $(\Gamma_{34}-1)m_3c^2$. A smaller $U_3$ in the total energy budget of Equation~\ref{eq:e_blast} leads to a higher $\Gamma_3$ during the reverse shock crossing, and consequently, the reverse shock is significantly weakened.  }
\end{itemize}

The lower panels show the thin-shell cases with varying upstream magnetizations; as the magnetization increases, the reverse shock becomes weaker and the crossing time decreases rapidly, eventually approaching the engine duration.

   \begin{figure*}
        \centering
        \includegraphics[width=\textwidth]{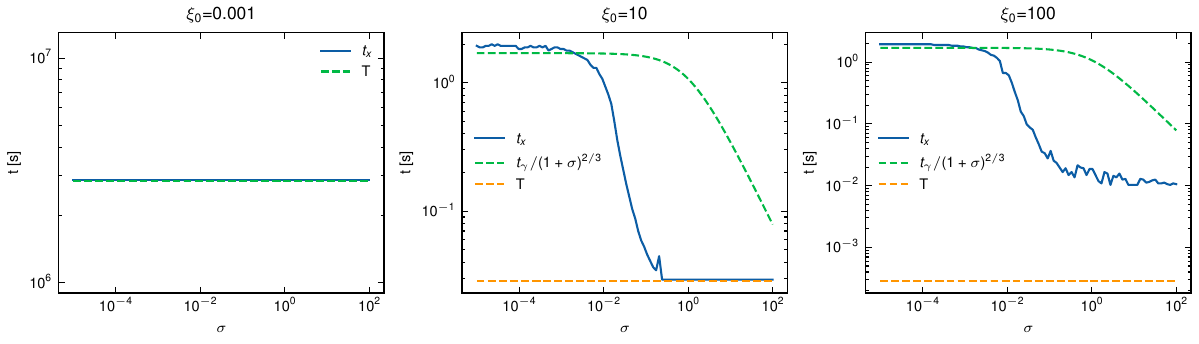}
        \caption{Numerical results for the reverse shock crossing time, $t_\times$, as a function of upstream magnetization, $\sigma$, for different shell thicknesses, based on Equation~\ref{eq:r-cross1} and \ref{eq:r-cross2}. The green dashed lines indicate the analytical scaling derived in \cite{zhang_kobayashi_05}, while the orange horizontal dashed lines represent the engine duration.}
        \label{fig:crossing}
    \end{figure*}

Figure~\ref{fig:crossing} shows our numerical results for the reverse shock crossing time, $t_\times$, as a function of upstream magnetization, $\sigma$, for different shell thicknesses, obtained by solving Equation~\ref{eq:r-cross1} and \ref{eq:r-cross2}. We remark that $t_\times$ does not strictly follow the analytical scaling derived in \cite{zhang_kobayashi_05}, where the deceleration radius coefficient was estimated as $C_\gamma \sim (1+\sigma)^{-1/3}$ and the shock crossing radius coefficient as $C_\Delta \sim (1+\sigma)^{-1/2}$. Under this assumption, the shock crossing time scales as $t_\times \sim t_\gamma \left(\frac{C_\Delta}{C_\gamma}\right)^4 \propto t_\gamma/(1+\sigma)^{2/3}$, where $t_\gamma \sim l/(2\Gamma_4^{8/3}c)$ is the deceleration time in the thin-shell regime for an unmagnetized upstream with Sedov length $l$.

The original derivation assumes that $C_\Delta = \left(1 - \frac{\Gamma_4 n_4}{\Gamma_3 n_3}\right)^{1/2}$ is effectively independent of $\Gamma_{43}$ and scales as $(1+\sigma)^{-1/2}$. This assumption holds for mildly and ultra-relativistic reverse shocks, as shown in Figure 3 of \cite{zhang_kobayashi_05}. However, based on our fully analytical shock jump conditions (Equation~\ref{eq:continuous}, \ref{eq:us}, and \ref{eq:u_down}), we find that $C_\Delta$ deviates from the $(1+\sigma)^{-1/2}$ scaling when $\Gamma_{43} - 1$ is small, which is typically the case during the early stage in thin-shell regime.

       \begin{figure}
        \centering
        \includegraphics[width=.45\textwidth]{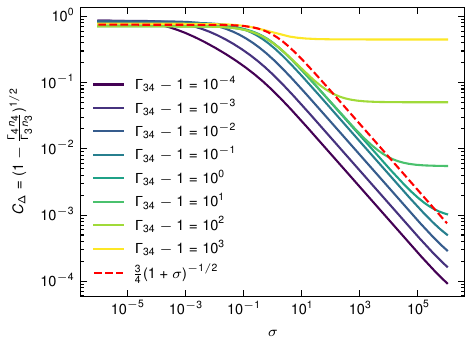}
        \caption{The reverse shock crossing radius coefficient $C_\Delta$ (See definition in Equation~39 of \cite{zhang_kobayashi_05}) as a function of upstream magnetization, $\sigma$, and reverse shock strength, $\Gamma_{34} - 1$. The upstream Lorentz factor is set to 2000. The red dashed line shows the analytical scaling derived in \cite{zhang_kobayashi_05}.}
        \label{fig:crossing-coef}
    \end{figure}

Figure~\ref{fig:crossing-coef} shows the shock crossing radius coefficient $C_\Delta$ as a function of magnetization $\sigma$ for different values of $\Gamma_{43} - 1$. Consistent with the result in \cite{zhang_kobayashi_05}, we find that for $\Gamma_{43} > 1.5$, $C_\Delta$ can be approximated as $\frac{3}{4}(1+\sigma)^{-1/2}$. However, in the thin-shell regime, where the reverse shock is weak and $\Gamma_{43} - 1 \ll 1$ at early times, $C_\Delta$ begins to deviate from this scaling well before $\sigma \sim 1$. As indicated in the figure, $C_\Delta$ starts to decrease significantly at lower values of $\sigma$, leading to the discrepancy between our numerical results and the analytical scaling in Figure~\ref{fig:crossing}. The late-time flattening observed in Figure~\ref{fig:crossing-coef} arises from the use of the exact expression for the relative Lorentz factor (Equation~\ref{eq:relative_gamma}), rather than the commonly used approximation $\Gamma_{43} \sim \frac{1}{2} \left( \frac{\Gamma_4}{\Gamma_3} + \frac{\Gamma_3}{\Gamma_4} \right)$, which may break down if either $\Gamma_4$ or $\Gamma_3$ does not satisfy the $\gg 1$ condition.

{ We remark that the reverse shock generation condition given in Equation~\ref{eq:rs-gen} remains valid in the energy conservation approach, since it is derived at the onset of the reverse shock crossing, when the blast wave is still infinitesimally thin. Under this circumstance, pressure balance and energy conservation are indeed equivalent.}

\subsubsection{Post Shock Crossing Dynamics}
{
    Once the reverse shock crosses the entire jet, it will gradually adjust to the Blandford–McKee self-similar solution~\citep{blandford_mckee_76}, while maintaining a constant number of shocked protons afterward. The forward shock will continue to propagate through the external medium. Several early works have addressed the blast wave dynamics after the reverse shock crossing \citep{Chiang1999, Piran1999, huang_etal_99,peer2012,Nava2013,Zhang2018book,Yuan2024}. \\

    {\centering (i) Forward shock\label{sec:forward}\\}

After the shock crossing, the shock heating and mass accumulation in region 3 cease, which we implement by turning off the corresponding terms in Equation~\ref{eq:mech-eqn}, where
\begin{eqnarray}
\frac{dm_3}{dt}&=&0,\\
\frac{dU_3}{dt}&=&-(\hat\gamma_{34}-1)U_3\bigg(\frac{2}{r}\frac{dr}{dt}+\frac{1}{\Delta_3^\prime}\frac{d\Delta_3^\prime}{dt}\bigg),\\
\frac{d\Delta_3^\prime}{dt} &=& c_{\rm s,3}\frac{dt^\prime}{dt},\\
c_{s,3}&=&c\sqrt{\frac{\hat\gamma_{34}(\hat\gamma_{34}-1)(\Gamma_{3}-1)}{1+\hat\gamma_{34}(\Gamma_{3}-1)}}.
\end{eqnarray}
One can then continue solving Equation~\ref{eq:mech-eqn} to obtain the forward shock dynamics. Indeed, if the adiabatic cooling in region 3 is ignored (cold blast wave assumption), this reduces to the forward shock dynamics given in Equation 8.60 of \cite{Zhang2018book}. It is worth noting that further simplification by setting $\hat{\gamma}_2 = 1$ (i.e., neglecting the pressure in the blast wave) does not recover the shock dynamics in \cite{huang_etal_99}. This discrepancy arises from different assumptions about shock heating. In \cite{huang_etal_99}, the authors adopt $dU_{\rm sh} = c^2 d[(\Gamma_2 - 1)m_2]$, whereas here we adopt $dU_{\rm sh} = c^2 (\Gamma_2 - 1)dm_2$.\\

}
    {\centering (ii) Reverse shock\\}
    
    After the reverse shock has crossed the shell, the assumption $\Gamma_2=\Gamma_3$ is no longer valid, as regions 2 and 3 subsequently evolve with generally different bulk Lorentz factors. The scaling laws shortly after the reverse shock crossing cannot be solved analytically. \cite{meszaros_rees_99} introduced a parameterized power-law decay behavior where
    \begin{eqnarray}
    &&u_3=\Gamma_3\beta_3 \propto r^{-g}\,,\\
    &&r \propto t^{1/(1+2g)},\quad u_3 \gg 1\,,\\
    &&r \propto t^{1/(1+g)},\quad u_3 \ll 1\,.
    \end{eqnarray}
    Hydrodynamic simulations from \cite{kobayashi_sari_00} indicate that $g\sim2.2$ and $\hat\gamma_{43}=4/3$ could adequately describe the dynamics in the thin shell regime, while $g\sim3$ and $\hat\gamma_{43}=4/3$ are good for the thick shell regime.

    Instead of adopting scaling prescriptions, we numerically calculate the post-shock crossing dynamics based on adiabatic expansion and mass conservation. Since the total number of shocked protons/electrons remains constant after shock crossing, we have
    \begin{eqnarray}
        n_3 r^2 \Delta_3^\prime \propto r^0\,,
    \end{eqnarray}
    thus
    \begin{eqnarray}
        n_3(r) = n_{3}(r_\times)\bigg(\frac{r}{r_\times}\bigg)^{-2}\bigg(\frac{\Delta_3^\prime}{\Delta_\times^\prime}\bigg)^{-1}\,,
    \end{eqnarray}
    where $r_\times$ is the radius where the reverse shock crosses the entire jet and $\Delta_\times^\prime$ is the corresponding co-moving shell width. By using the adiabatic expansion approximation,
    \begin{eqnarray}
        p_3\propto \rho_3^{\hat{\gamma}_{43}} \propto n_3^{\hat{\gamma}_{43}}\,,
    \end{eqnarray}
    we get
    \begin{eqnarray}
        p_3(r) = p_{3}(r_\times)\bigg(\frac{r}{r_\times}\bigg)^{-2\hat{\gamma}_{43}}\bigg(\frac{\Delta_3^\prime}{\Delta_\times^\prime}\bigg)^{-\hat{\gamma}_{43}}\,.
    \end{eqnarray}
    Then, the ``temperature" of the electron in region 3 ($\Gamma_{43}-1$), can be expressed as
    \begin{eqnarray}
        \Gamma_{43}-1 \propto \frac{p_3}{n_3}\propto \bigg(\frac{r}{r_\times}\bigg)^{2(1-\hat{\gamma}_{43})}\bigg(\frac{\Delta_3^\prime}{\Delta_\times^\prime}\bigg)^{1-\hat{\gamma}_{43}}\,.
    \end{eqnarray}
     This treatment automatically recovers all the regimes with correct $\Delta_3^\prime$ spreading. The shell starts to expand adiabatically after the shock crossing radius, $r_\times$, 
    \begin{eqnarray}
        \Delta_3^\prime \propto {\rm c_{s,3}}\frac{r}{\Gamma_3} \propto\left\{
        \begin{aligned}
            &\sqrt{\hat\gamma_{43}\frac{p_3}{n_3}}\frac{r}{\Gamma_3}\propto r^{\frac{2(2+g-\hat\gamma_{43})}{\hat\gamma_{43}+1}},\quad &\Gamma_{43}-1\ll1 \\
            &\sqrt{\hat\gamma_{43}-1}\frac{r}{\Gamma_3} \propto r^{1+g},\quad &\Gamma_{43}-1\gg1
        \end{aligned}
        \right.\,.
    \end{eqnarray}

    In the thin shell regime, since the reverse shock is Newtonian ($\Gamma_{43}-1\ll1$), we have
    \begin{eqnarray}
        n_3 &\propto& r^{-\frac{2(3+g)}{\hat\gamma_{43}+1}}\,,\\
        p_3 &\propto& r^{-\frac{2(3+g)\hat\gamma_{43}}{\hat\gamma_{43}+1}}\,,\\
        \frac{p_3}{n_3} &\propto& r^{-\frac{2(3+g)(\hat\gamma_{43}-1)}{\hat\gamma_{43}+1}}\,.
    \end{eqnarray}
    This is reduced to \cite{kobayashi_sari_00} if we take $\hat\gamma_{43}=4/3$, where they calculate the sub-relativistic reverse shock case while assuming that the electron is still hot enough to support $\hat\gamma_{43}=4/3$. 

    In the thick shell regime ($\Gamma_{43}-1\gg1$), the reverse shock becomes relativistic. During the adiabatic expansion, we have
    \begin{eqnarray}
        n_3 &\propto& r^{-(3+g)}\,,\\
        p_3 &\propto& r^{-(3+g)\hat\gamma_{43}}\propto r^{-\frac{4(3+g)}{3}}\,,\\
        \frac{p_3}{n_3} &\propto& r^{(3+g)(1-\hat\gamma_{43})} \propto r^{-\frac{3+g}{3}}\,.
    \end{eqnarray}
    By taking $g=7/2$, one may recover the scalings in \cite{kobayashi_sari_00}.

    \begin{figure}
        \centering
        \includegraphics[width=.45\textwidth]{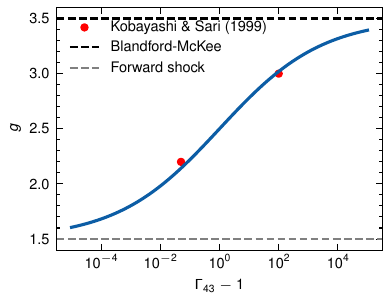}
        \caption{The post reverse shock crossing $g$ parameter as a function of the reverse shock strength $\Gamma_{43}$ }
        \label{fig:g}
    \end{figure}

    Numerically, for the post reverse shock crossing $g$ profile, we adopt a $\Gamma_{43}$-dependent $g$ parameter that approaches $3/2$ as $\Gamma_{43}-1\rightarrow 0$ (i.e. $\Gamma_3$ follows the same scaling as $\Gamma_2$ in ISM post crossing) and approaches $7/2$ as $\Gamma_{43}-1\rightarrow \infty$ \citep[i.e.,][]{blandford_mckee_76}.  Figure~\ref{fig:g} shows our prescription of the $\Gamma_{43}$-dependent $g$ parameter. The red dots show hydrodynamical simulation results from \cite{kobayashi_sari_00}, where they obtained $\Gamma_{43}-1\sim0.05$, $g\sim2.2$ and $\Gamma_{43}-1\sim100$, $g\sim3$.

\subsection{Jet spreading}\label{sec:spreading}
The lateral expansion of conical jets can steepen the power-law decay index of the late-time afterglow light curve \citep{rhoads_99,sari_etal_99,granot_piran_12,duffell_etal_18, lu_etal_20}. It has been suggested that lateral expansion becomes important after the jet break, with the jet opening angle evolving as  
\begin{eqnarray}
    \frac{d\theta_j}{dt} \sim \frac{1}{\Gamma} \frac{\dot{r}}{r} \propto \frac{1}{\Gamma}.
\end{eqnarray}
However, hydrodynamical simulations indicate that this $\propto 1/\Gamma$ prescription overestimates the lateral expansion speed of conical jets, suggesting that significant lateral spreading does not occur until the jet becomes mildly relativistic. \citet{granot_piran_12} therefore proposed a revised prescription:
\begin{eqnarray}
     \frac{d\theta_j}{dt} \sim \frac{1}{\Gamma^{1+a}\theta_j^a} \frac{\dot{r}}{r},
\end{eqnarray}
where $a=1$ in the early stage when the shock velocity vector is nearly aligned with the shock front normal, and $a=0$ in the late stage when lateral spreading becomes non-negligible. This new formulation agrees better with hydrodynamical simulations. 

Although these prescriptions were developed for top-hat jets, more comprehensive hydrodynamical modeling is required for structured jets to properly account for pressure gradients across the jet profile, beyond the approximation of a rarefaction wave expanding into vacuum \citep{lu_etal_20}. Nonetheless, a structured jet can be decomposed into a series of concentric top-hat jets, each evolving according to the rarefaction wave expansion model \citep{ryan_20}.

Here, we introduce a smooth transition function that captures both the early and late phases of jet lateral expansion:
\begin{eqnarray}
\frac{d\theta_j}{dt} &=& F(u) \frac{1}{2\Gamma} \frac{\dot{r}}{r}, \\
F(u) &=& (1+Qu\theta_s)^{-1},
\end{eqnarray}
where $u = \Gamma \beta$.

\begin{figure}
    \centering
    \includegraphics[width=.45\textwidth]{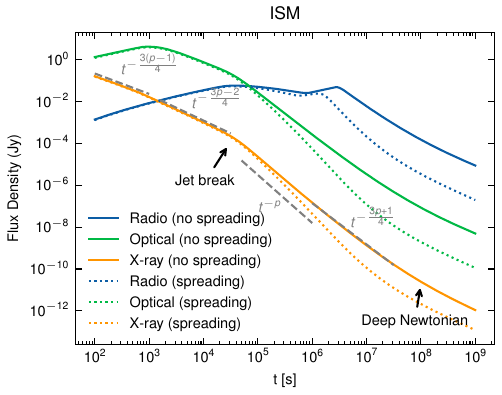}
    \includegraphics[width=.45\textwidth]{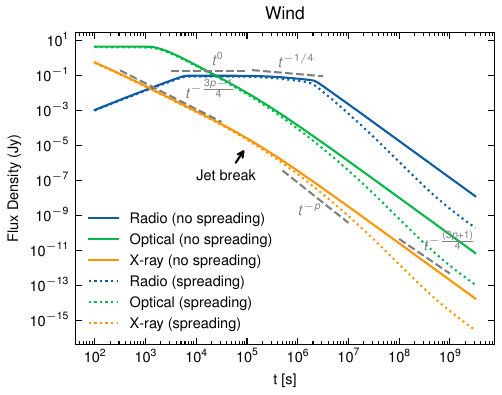}
    \caption{Light curves in the radio, optical, and X-ray bands for spreading and non-spreading top-hat jets observed on-axis. The upper panel displays light curves for a jet propagating into the interstellar medium (ISM), while the bottom panel shows the case for a jet propagating into a stellar wind.}\label{fig:spreading}
\end{figure}

Figure~\ref{fig:spreading} shows examples of top-hat jet spreading with an opening angle of $10^\circ$ in both ISM and stellar wind environments, across the radio, optical, and X-ray bands. The figure demonstrates that, under this prescription, the jet begins to spread significantly later than the jet break. The post-expansion temporal decay can be approximated as $t^{-p}$ before the system enters the deep Newtonian regime.

\section{Radiation}\label{sec:radiation}
All radiations are first calculated in the fluid co-moving frame and then Doppler-transformed to the observer frame. The advantage of performing the radiation calculation in the co-moving frame is that, for an axisymmetric structured jet, the radiation becomes $\phi$-independent. As a result, the intrinsic emission only needs to be computed on a 2D grid  of ($\theta$, $t$) rather than a full 3D grid of ($\phi$, $\theta$, $t$), even for an off-axis observer. This significantly reduces the computational cost. The effects of observer geometry can then be incorporated separately through grid-based Doppler transformation. All discussions in this section refer to quantities defined in the fluid co-moving frame.

\subsection{Minimal injection Lorentz factor}\label{sec:gamma_m}
    
In the context of relativistic shocks, the dissipation of kinetic energy at the shock front leads to a sharp increase in internal energy and pressure in the downstream region, accelerating particles to relativistic energies.
The comoving internal energy density behind the shock can be obtained from
\begin{equation}
    \mathcal{U}=U/V^\prime\,,
\end{equation}
where $V^\prime$ is the comoving volume of the blast wave\footnote{Note that this is numerically slightly different from the conventional $\mathcal{U}=(\Gamma-1)n_pm_pc^2$, where $n_p$ is the downstream proton number density, which can be obtained from the shock jump conditions (Equation~\ref{eq:continuous})}.

In the shock downstream, a fraction of internal energy is typically transferred to non-thermal electrons with a mean energy density given by
\begin{eqnarray}
\label{eq:U_electrons}
   (\bar\gamma -1)n_em_ec^2 =\epsilon_e\mathcal{U},
\end{eqnarray}
where $\bar\gamma$ is the average Lorentz factor of the accelerated electrons. The shocked electron energy distribution is typically assumed to follow a power-law distribution with index $p$, 
\begin{eqnarray}
\label{eq:e_distribution}
    \frac{dN_e}{d\gamma} \propto (\gamma-1)^{-p}, \quad  &\gamma_m \leq \gamma \leq \gamma_M,
\end{eqnarray}
where $\gamma_m$ and $\gamma_M$ (See definition in Section~\ref{sec:gamma_max}) are the minimum and maximum Lorentz factor of electrons, respectively. The minimum Lorentz factor, $\gamma_m$, can then be determined from Equation~\ref{eq:U_electrons} and \ref{eq:e_distribution}, see, e.g. \citealt{sari_etal_98,panaitescu_02,Dai2001, Zhang2018book}, but with $(+1)$ added at the end:
\begin{eqnarray}
    \gamma_m=\left\{
    \begin{aligned}
    &\frac{p-2}{p-1}\frac{\epsilon_e}{\xi_e}(\Gamma-1)\frac{m_p}{m_e}+1, \quad &p>2\\
    &\ln^{-1}(\frac{\gamma_M}{\gamma_m})\frac{\epsilon_e}{\xi_e}(\Gamma-1)\frac{m_p}{m_e}+1, \quad &p=2\\
    &\bigg(\frac{p-2}{p-1}\frac{\epsilon_e}{\xi_e}(\Gamma-1)\frac{m_p}{m_e}\gamma_M^{p-2}\bigg)^{1/(p-1)}+1,\quad &p<2
    \end{aligned}
    \right.
\end{eqnarray}
where $\xi_e={n_e}/{n_p}$ is the fraction of electrons that are accelerated, by using
\begin{eqnarray}
    \bar\gamma_m-1=\frac{\int_{\gamma_m}^{\gamma_M}(\gamma-1)\frac{dN_e}{d\gamma}d\gamma}{\int_{\gamma_m}^{\gamma_M}\frac{dN_e}{d\gamma}d\gamma}\,.
\end{eqnarray}

The $+1$ correction on $\gamma_m$ comes from that the kinetic electron energy $\epsilon_e\mathcal{U} \propto \gamma_m-1$ (not $\propto \gamma_m$). This is valid in both relativistic and non-relativistic regimes and guarantees that $\gamma_m>1$. However, at low electron energies, synchrotron radiation transitions into the cyclotron regime, where emission is dominated by the fundamental harmonic, and relativistic broadening becomes negligible. Since the electron population follows a power-law distribution in energy, a portion of the electrons may lie close to or within the cyclotron regime, where synchrotron emission is inefficient. To account for this, a correction factor, $f_{\rm syn}$, is required to suppress the synchrotron flux at lower energies, defined as
\begin{eqnarray}
f_{\rm syn}=\bigg(\frac{\gamma_m-1}{\gamma_m}\bigg)^\kappa,
\end{eqnarray}
where $\kappa$ is a $p$-dependent power-law index. If one assumes that the electron energy distribution follows a power-law distribution with index $p$ down to the non-relativistic regime \citep{huang_03,frail_etal_04}, then $\kappa=p-1$. However, \cite{sironi_13} remarked that this violates the Fermi acceleration theory in the low Lorentz factor regime and indeed the momentum of the electron follows the power-law distribution,
\begin{eqnarray}
    \frac{dN_e}{dy}\propto (y^2+2y)^{-(p+1)/2}(y+1) \propto \gamma(\gamma^2+1)^{-(p+1)/2}\,,
\end{eqnarray}
where $y=\gamma-1$. With this distribution, one can find
\begin{eqnarray}
    \kappa=\left\{
    \begin{aligned}
        &0,\quad &2<p<3 \\
        &({p-1})/{2},\quad &p>3
    \end{aligned}
    \right.\,.
\end{eqnarray}

We remark that, without the $+1$ correction on $\gamma_m$ in the non-relativistic regime, the flux in the deep Newtonian regime will be underestimated. Furthermore, with the $+1$ correction on $\gamma_m$ but without the $f_{\rm syn}$ correction, the flux in the deep Newtonian regime will be overestimated (overflattened).

\subsection{Synchrotron \& IC Cooling}
As electrons emit synchrotron photons, they lose energy at a rate governed by the magnetic field strength, the initial electron velocities, and additional scattering processes (Inverse Compton). For synchrotron and IC cooling, in the blast wave co-moving frame, one can write
\begin{eqnarray}
\frac{d\gamma}{dt^\prime}m_ec^2=-(P_{\rm syn}+P_{\rm IC})=-\frac{4}{3}\sigma_Tc\gamma^2\beta^2\frac{B^{\prime2}}{8\pi}(1+\tilde{Y}),
\end{eqnarray}
where $\sigma_T$ is the Thomson scattering cross-section and $\tilde{Y}(\gamma)$ is the Compton Y-parameter. The characteristic comoving timescale for this energy loss, $t'_c$, is given by,
\begin{eqnarray}
    t_c^\prime=\bigg|\frac{\gamma_c}{\dot{\gamma_c}}\bigg|=\frac{6\pi m_ec^2\gamma_c}{\sigma_Tc(\gamma_c^2-1)B^{\prime2}(1+\tilde{Y})}.
\end{eqnarray}

Then the typical cooling Lorentz factor, $\gamma_c$, beyond which the electrons are cooled by synchrotron and IC radiation can be expressed as
\begin{eqnarray}
    \gamma_c&=&\frac{1}{2}\bigg(\bar{\gamma}_c+\sqrt{\bar{\gamma}^2_c+4}\bigg)\,,\\
    \bar{\gamma}_c&=&\frac{6\pi m_ec}{\sigma_TB^{\prime2}(1+\tilde{Y})t^\prime}\,.
\end{eqnarray}
This solution (where $P_{\rm syn}\propto \gamma^2\beta^2$ instead of just $\propto \gamma^2$ which is only valid in the relativistic regime) applies to non-relativistic regimes as well, where $t^\prime \gg \frac{\sigma_TB^{\prime2}(1+\tilde{Y})}{6\pi m_ec}$.

\subsection{Maximum Lorentz Factor}\label{sec:gamma_max}
The maximum particle energy from the Fermi acceleration process can be estimated by balancing acceleration and cooling within the dynamical time scale. For electrons, the comoving acceleration time scale can be written as
\begin{eqnarray}
    t^\prime_{\rm acc} \sim \frac{\gamma m_ec}{eB^\prime}\,,
\end{eqnarray}
where $B^\prime$ is the co-moving magnetic field in the downstream. Then by setting $t^\prime_c=t^\prime_{\rm acc}$, one gets
\begin{eqnarray}
    \gamma_M = \bigg(\frac{6\pi e}{\sigma_TB^\prime(1+\tilde{Y})}\bigg)^{1/2}\,.
\end{eqnarray}
Beyond this Lorentz factor, the electron number density decays exponentially as $\propto e^{-{\gamma^2}/{\gamma_M^2}}$. 

\subsection{Synchrotron Self-absorption}
The self-absorption frequency, $\nu_a$, defines the characteristic frequency below which synchrotron photons are self-absorbed. In the self-absorption regime, the spectral index resembles the Rayleigh-Jeans regime of a blackbody spectrum. Therefore, $\nu_a$ can be derived as the point of intersection between the two spectra,
\begin{eqnarray}
    I_{\rm \nu}^{\rm bb}(\nu_a)&=&I_{\nu}^{\rm syn}(\nu_a)\sim 2kT\frac{\nu_a^2}{c^2}\,,\\
    kT&=&(\gamma_{p}-1)m_ec^2,
\end{eqnarray}
where $I_\nu$ is the specific synchrotron intensity (see section~\ref{sec:syn_rad}) and $kT$ is the effective blackbody temperature. Then one can solve the self-absorption frequency for weak absorption $\nu_{a}<{\rm min}(\nu_m,\nu_c)$ (intersection in the $1/3$ segment), intermediate absorption $\nu_m<\nu_a<\nu_c$ (intersection in the $-(p-1)/2$ segment) and strong absorption $\nu_c<\nu_a$ (electron pile-up around $\gamma_a$)
\begin{eqnarray}
    \nu_a =\left\{
    \begin{aligned}
    &\bigg(\frac{I_{\rm \nu,max}}{2(\gamma_p-1)m_e\nu_{p}^{1/3}}\bigg)^{\frac{3}{5}},\quad &\nu_{a}< {\rm min}(\nu_m,\nu_c)\\
    &\bigg(\frac{I_{\rm \nu,max}}{2m_e}\sqrt{\frac{3eB^\prime}{2\pi m_ec}}\bigg)^{\frac{2}{p+4}}\nu_m^{\frac{p-1}{p+4}},\quad &\nu_m<\nu_a<\nu_c\\
    &\bigg(\frac{I_{\rm \nu,max}}{2m_e}\sqrt{\frac{3eB^\prime}{2\pi m_ec}}\bigg)^{\frac{2}{5}},\quad &\nu_c<\nu_a
    \end{aligned}
    \right.
\end{eqnarray}
where $I_{\rm \nu, max}$ is the peak specific intensity of $I_\nu$, which writes
\begin{eqnarray}
    I_{\rm\nu,max}=\bar P_{\rm syn, max}\frac{f_{\rm syn}N_{\rm p, tot}\xi_e}{4\pi \delta^\prime}=0.92\frac{\pi}{4}\frac{\sqrt{3}B^\prime e^3}{m_ec^2}\frac{f_{\rm syn}\Sigma_{\rm p, tot}\xi_e}{4\pi}\,,
\end{eqnarray}
where $\Sigma_{\rm p, tot}=\frac{N_{\rm p, tot}}{4\pi r^2}$ is the column density of the shocked proton and $N_{\rm p, tot}$ is the total number of shocked protons. The coefficient $0.92$ arises from the maximum value of the single-electron synchrotron spectrum function $\mathcal{F}(x)$ \citep{Rybicki1979}, and the factor $\frac{\pi}{4}$ comes from the average pitch angle $\langle\sin\alpha\rangle$. { The blackbody method adopted here is shown to be consistent with the optical depth method \citep{Shen2009} implemented in {\tt PyFRS} \citep{Lei2016} and {\tt PyBlastAfterglow} \citep{Nedora2024}.}

\subsection{Electron Spectrum}\label{sec:e_spectrum}
With a given order of the characteristic Lorentz factors, $\gamma_m$, $\gamma_c$ and $\gamma_a$, and total synchrotron electron number $N_{\rm e, tot}=f_{\rm syn}N_{\rm p, tot}\xi_e$, the electron number spectrum with synchrotron cooling and self-absorption can be approximately expressed as a broken power-law for six different regimes~\citep{sari_etal_98, granot_sari_02, Kobayashi2004, gao_13b, Zhang2018book}.\\

(I) $\gamma_a<\gamma_m<\gamma_c$ (slow cooling, weak absorption) and (II) $\gamma_m<\gamma_a<\gamma_c$ (slow cooling, weak absorption),
\begin{eqnarray}
        \frac{dN_e}{d\gamma}=N_{\rm e, tot}\left\{
        \begin{aligned}
            &(p-1)\gamma_m^{p-1}\gamma^{-p},\quad &\gamma_m<\gamma<\gamma_c\\
            &(p-1)\gamma_m^{p-1}\gamma_c\gamma^{-p-1},\quad &\gamma_c<\gamma
        \end{aligned}
        \right.\,.
\end{eqnarray}
\\

(III) $\gamma_a<\gamma_c<\gamma_m$ (fast cooling, weak absorption), 
\begin{eqnarray}
        \frac{dN_e}{d\gamma}=N_{\rm e, tot}\left\{
        \begin{aligned}
            &\gamma_c\gamma^{-2},\quad &\gamma_c<\gamma<\gamma_m\\
            &\gamma_m^{p-1}\gamma_c\gamma^{-p-1},\quad &\gamma_m<\gamma
        \end{aligned}
        \right.\,.
\end{eqnarray}
\\

(IV) $\gamma_c<\gamma_a<\gamma_m$ (fast cooling, weak absorption),
\begin{eqnarray}
        \frac{dN_e}{d\gamma}=N_{\rm e, tot}\left\{
        \begin{aligned}
            &3\gamma_a^{-3}\gamma^{2},\quad &\gamma<\gamma_a\\
            &\gamma_c\gamma^{-2},\quad &\gamma_a<\gamma<\gamma_m\\
            &\gamma_m^{p-1}\gamma_c\gamma^{-p-1},\quad &\gamma_m<\gamma
        \end{aligned}
        \right.\,.
\end{eqnarray}
\\

(V) $\gamma_m<\gamma_c<\gamma_a$ (slow cooling, strong absorption),
\begin{eqnarray}
        \frac{dN_e}{d\gamma}=N_{\rm e, tot}\left\{
        \begin{aligned}
            &3\gamma_a^{-3}\gamma^{2},\quad &\gamma<\gamma_a\\
            &(p-1)\gamma_m^{p-1}\gamma_c\gamma^{-p-1},\quad &\gamma_a<\gamma
        \end{aligned}
        \right.\,.
\end{eqnarray}
\\

(VI) $\gamma_c<\gamma_m<\gamma_a$ (fast cooling, strong absorption),
\begin{eqnarray}
        \frac{dN_e}{d\gamma}=N_{\rm e, tot}\left\{
        \begin{aligned}
            &3\gamma_a^{-3}\gamma^{2},\quad &\gamma<\gamma_a\\
            &\gamma_m^{p-1}\gamma_c\gamma^{-p-1},\quad &\gamma_a<\gamma
        \end{aligned}
        \right.\,.
\end{eqnarray}
\\

\subsection{Synchrotron spectrum}\label{sec:syn_rad}
\begin{figure*}
        \centering
        \includegraphics[width=0.33\textwidth]{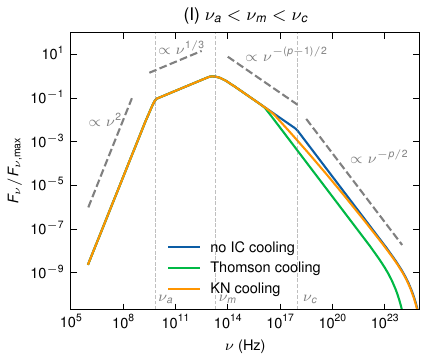}
        \includegraphics[width=0.33\textwidth]{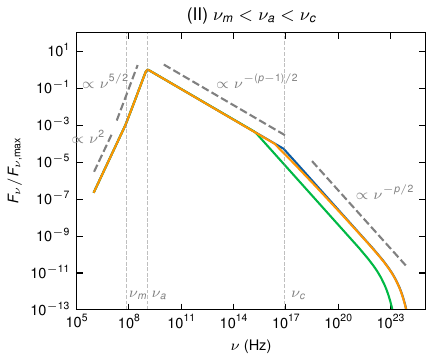}
        \includegraphics[width=0.33\textwidth]{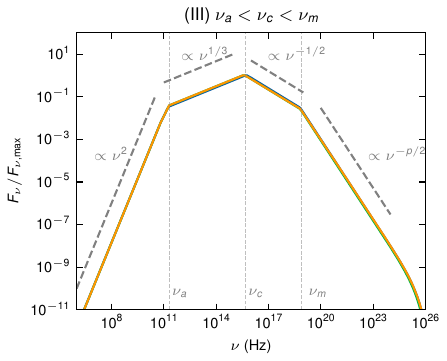}\\
        \includegraphics[width=0.33\textwidth]{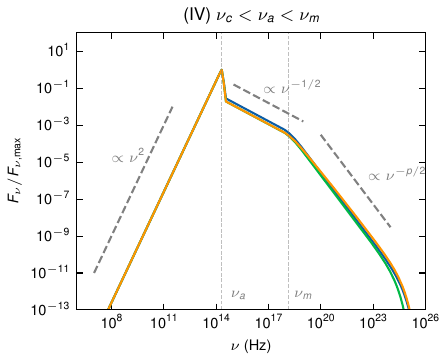}
        \includegraphics[width=0.33\textwidth]{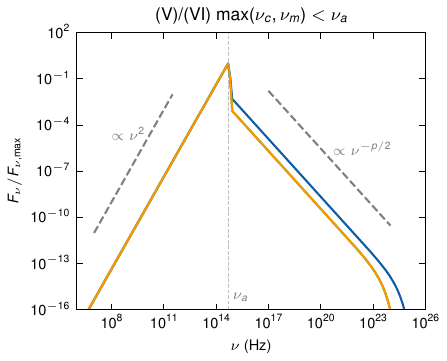}
        \caption{Observed synchrotron spectra from a narrow top‐hat jet viewed on‐axis, illustrating five characteristic spectral shapes. Blue curves show the spectra without any inverse Compton cooling; green curves include inverse Compton cooling but omit Klein–Nishina corrections; orange curves include inverse Compton cooling with full Klein–Nishina corrections.}
        \label{fig:spectrum}
\end{figure*}

Analogous to the electron energy distribution, the synchrotron radiation spectrum can be approximated by a multi-segment broken power-law function. The physical scenario is determined by the specific ordering of the characteristic frequencies: the self-absorption frequency ($\nu_a$), the characteristic synchrotron frequency ($\nu_m$), and the cooling frequency ($\nu_c$). For a given characteristic Lorentz factor, the  corresponding characteristic synchrotron emission frequency can be calculated from
\begin{eqnarray}
    \nu = \frac{3eB^\prime\gamma^2}{4\pi m_e c}\,.
\end{eqnarray}

Here, we list the functional forms for the various regimes:\\
(I) $\nu_a<\nu_m<\nu_c$ (slow cooling, weak absorption),
\begin{eqnarray}
    I_\nu=I_{\rm\nu,max}e^{-\nu/\nu_M}\left\{
    \begin{aligned}
        &\bigg(\frac{\nu_a}{\nu_m}\bigg)^{1/3}\bigg(\frac{\nu}{\nu_a}\bigg)^{2},\quad &\nu<\nu_a\\
        &\bigg(\frac{\nu}{\nu_m}\bigg)^{1/3},\quad &\nu_a<\nu<\nu_m\\
        &\bigg(\frac{\nu}{\nu_m}\bigg)^{-(p-1)/2},\quad &\nu_m<\nu<\nu_c\\
        &\bigg(\frac{\nu_c}{\nu_m}\bigg)^{-(p-1)/2}\bigg(\frac{\nu}{\nu_c}\bigg)^{-p/2},\quad &\nu_c<\nu\\
    \end{aligned}
    \right.\,.
\end{eqnarray}
\\
(II) $\nu_m<\nu_a<\nu_c$ (slow cooling, weak absorption),
\begin{eqnarray}
    I_\nu=I_{\rm\nu,max}e^{-\nu/\nu_M}\left\{
    \begin{aligned}
        &\bigg(\frac{\nu_m}{\nu_a}\bigg)^{(p+4)/2}\bigg(\frac{\nu}{\nu_m}\bigg)^{2},\quad &\nu<\nu_m\\
        &\bigg(\frac{\nu_a}{\nu_m}\bigg)^{-(p-1)/2}\bigg(\frac{\nu}{\nu_a}\bigg)^{5/2},\quad &\nu_m<\nu<\nu_a\\
        &\bigg(\frac{\nu}{\nu_m}\bigg)^{-(p-1)/2},\quad &\nu_a<\nu<\nu_c\\
        &\bigg(\frac{\nu_c}{\nu_m}\bigg)^{-(p-1)/2}\bigg(\frac{\nu}{\nu_c}\bigg)^{-p/2},\quad &\nu_c<\nu\\
    \end{aligned}
    \right.\,.
\end{eqnarray}
\\
(III) $\nu_a<\nu_c<\nu_m$ (fast cooling, weak absorption), 
\begin{eqnarray}
    I_\nu=I_{\rm\nu,max}e^{-\nu/\nu_M}\left\{
    \begin{aligned}
        &\bigg(\frac{\nu_a}{\nu_c}\bigg)^{1/3}\bigg(\frac{\nu}{\nu_a}\bigg)^{2},\quad &\nu<\nu_a\\
        &\bigg(\frac{\nu}{\nu_c}\bigg)^{1/3},\quad &\nu_a<\nu<\nu_c\\
        &\bigg(\frac{\nu}{\nu_c}\bigg)^{-1/2},\quad &\nu_c<\nu<\nu_m\\
        &\bigg(\frac{\nu_m}{\nu_c}\bigg)^{-1/2}\bigg(\frac{\nu}{\nu_m}\bigg)^{-p/2},\quad &\nu_m<\nu\\
    \end{aligned}
    \right.\,.
\end{eqnarray}
\\
(IV) $\nu_c<\nu_a<\nu_m$ (fast cooling, strong absorption),

\begin{eqnarray}
    I_\nu=I_{\rm\nu,max}e^{-\nu/\nu_M}\left\{
    \begin{aligned}
        &\bigg(\frac{\nu}{\nu_a}\bigg)^{2},\quad &\nu<\nu_a\\
        &\mathcal{R}_4\bigg(\frac{\nu}{\nu_a}\bigg)^{-1/2},\quad &\nu_a<\nu<\nu_m\\
        &\mathcal{R}_4\bigg(\frac{\nu_m}{\nu_a}\bigg)^{-1/2}\bigg(\frac{\nu}{\nu_m}\bigg)^{-p/2},\quad &\nu_m<\nu\\
    \end{aligned}
    \right.\,.
\end{eqnarray}
where $\mathcal{R}_4=\frac{\gamma_c}{3\gamma_a}$.
\\
(V) $\nu_m<\nu_c<\nu_a$ (slow cooling, strong absorption),
 \begin{eqnarray}
    I_\nu=I_{\rm\nu,max}e^{-\nu/\nu_M}\left\{
    \begin{aligned}
        &\bigg(\frac{\nu}{\nu_a}\bigg)^{2},\quad &\nu<\nu_a\\
        &\mathcal{R}_5\bigg(\frac{\nu}{\nu_a}\bigg)^{-p/2},\quad &\nu_a<\nu\\
    \end{aligned}
    \right.\,.
\end{eqnarray}
where $\mathcal{R}_5=(p-1)\frac{\gamma_c}{3\gamma_a}\bigg(\frac{\gamma_m}{\gamma_a}\bigg)^{p-1}$.
\\
(VI) $\nu_c<\nu_m<\nu_a$ (fast cooling, strong absorption),
\begin{eqnarray}
    I_\nu=I_{\rm\nu,max}e^{-\nu/\nu_M}\left\{
    \begin{aligned}
        &\bigg(\frac{\nu}{\nu_a}\bigg)^{2},\quad &\nu<\nu_a\\
        &\mathcal{R}_6\bigg(\frac{\nu}{\nu_a}\bigg)^{-p/2},\quad &\nu_a<\nu\\
    \end{aligned}
    \right.\,.
\end{eqnarray}
where $\mathcal{R}_6=\frac{\gamma_c}{3\gamma_a}\bigg(\frac{\gamma_m}{\gamma_a}\bigg)^{p-1}$.

Figure~\ref{fig:spectrum} shows an example of on-axis observed synchrotron spectra for narrow opening angle top-hat jets (thus mimicking the shape of the intrinsic spectrum) with afterglow parameters in the six different regimes.

\subsection{Inverse Compton Cooling}\label{sec:ic}   
High-energy electrons can upscatter low-energy photons ($h\nu^\prime\ll m_ec^2$) to higher energies via the inverse Compton (IC) process. During the IC process, a synchrotron electron transfers energy to a photon, thereby undergoing further cooling compared to just synchrotron radiation. The extra cooling is usually characterized by the so-called $\tilde{Y}$ parameter that is used to describe the relative cooling efficiency compared to synchrotron cooling. In the Thompson regime, where $h\nu^\prime\ll m_ec^2$, IC cooling follows the same scaling of synchrotron cooling as $\propto \gamma^2\beta^2$, thus, $\tilde{Y}$ is a $\gamma$-independent constant. However, as the photon energy becomes larger ($h\nu^\prime\gg m_ec^2$), the IC process becomes less efficient so the IC cooling no longer follows $\propto \gamma^2\beta^2$. Therefore, $\tilde{Y}$ becomes $\gamma$-dependent, making more sub-segments on the electron and synchrotron spectra. With IC cooling, the cooled electron/synchrotron spectra can be expressed as \citep{Nakar2009,Wang2010,Jacovich2021, McCarthy2024},
\begin{eqnarray}
        \frac{dN^{\rm IC}_e}{d\gamma}(\gamma)=\frac{dN_e}{d\gamma}(\gamma)\left\{
        \begin{aligned}
        &1,\quad &\gamma<\gamma_c\\
        &\frac{1+\tilde{Y}(\gamma_c)}{1+\tilde{Y}(\gamma)},\quad &\gamma>\gamma_c
        \end{aligned}
        \right.\,,\label{eq:N_ic}
\end{eqnarray}
and 
\begin{eqnarray}
        I_\nu^{\rm syn, IC}(\nu)=I_\nu^{\rm syn}(\nu)\left\{
        \begin{aligned}
        &1,\quad &\nu<\nu_c\\
        &\frac{1+\tilde{Y}(\nu_c)}{1+\tilde{Y}(\nu)},\quad &\nu>\nu_c
        \end{aligned}
        \right.\,,\label{eq:syn_ic}
\end{eqnarray}
where below $\gamma_c$ ($\nu_c$), the electron (synchrotron) spectrum remains unchanged, while above $\gamma_c$ ($\nu_c$), a correction factor $\propto 1/(1+\tilde{Y})$ has been applied.

For a given characteristic frequency $\nu_m$, $\nu_c$ or $\nu_a$, there is a typical electron Lorentz factor beyond which IC scattering shifts from the Thompson regime to the Compton regime, given by,
\begin{eqnarray}
    \hat\gamma_i=\frac{m_ec^2}{h\nu_i}\,.
\end{eqnarray}
If one assumes a step-like Compton cross-section, beyond $\hat\gamma_i$, then the cross-section of upscattering the corresponding photon drops to zero and one can approximate $\tilde{Y}(\gamma)$ as \footnote{Note that such an approximation works only for weak Klein-Nishina regime where $\hat{\gamma}_m>\gamma_m$.}
\begin{eqnarray}
    \tilde{Y}(\gamma)=\left\{
        \begin{aligned}
            &Y_T,\quad &\gamma<\hat\gamma_m\\
            &Y_T\bigg(\frac{\gamma}{\hat\gamma_m}\bigg)^{-1/2},\quad &\hat\gamma_m<\gamma<\hat\gamma_c\\
            &Y_T\bigg(\frac{\gamma}{\gamma_c}\bigg)^{-4/3}\bigg(\frac{\hat\gamma_m}{\hat\gamma_c}\bigg)^{1/2},\quad &\hat\gamma_c<\gamma
        \end{aligned}
        \right.\,,
\end{eqnarray}

 \begin{eqnarray}
        \tilde{Y}(\nu)=\left\{
        \begin{aligned}
            &Y_T,\quad &\nu<\hat\nu_m\\
            &Y_T\bigg(\frac{\nu}{\hat\nu_m}\bigg)^{-1/4},\quad &\hat\nu_m<\nu<\hat\nu_c\\
            &Y_T\bigg(\frac{\nu}{\nu_c}\bigg)^{-2/3}\bigg(\frac{\hat\nu_m}{\hat\nu_c}\bigg)^{1/4},\quad &\hat\nu_c<\nu
        \end{aligned}
        \right.\,,
    \end{eqnarray}

for fast cooling, and

\begin{eqnarray}
        \tilde{Y}(\gamma)=\left\{
        \begin{aligned}
            &Y_T,\quad &\gamma<\hat\gamma_c\\
            &Y_T\bigg(\frac{\gamma}{\hat\gamma_c}\bigg)^{(p-3)/2},\quad &\hat\gamma_c<\gamma<\hat\gamma_m\\
            &Y_T\bigg(\frac{\gamma}{\gamma_m}\bigg)^{-4/3}\bigg(\frac{\hat\gamma_m}{\hat\gamma_c}\bigg)^{(p-3)/2},\quad &\hat\gamma_c<\gamma
        \end{aligned}
        \right.\,,
\end{eqnarray}

\begin{eqnarray}
        \tilde{Y}(\nu)=\left\{
        \begin{aligned}
            &Y_T,\quad &\nu<\hat\nu_c\\
            &Y_T\bigg(\frac{\nu}{\hat\nu_c}\bigg)^{(p-3)/4},\quad &\hat\nu_c<\nu<\hat\nu_m\\
            &Y_T\bigg(\frac{\nu}{\nu_m}\bigg)^{-2/3}\bigg(\frac{\hat\nu_m}{\hat\nu_c}\bigg)^{(p-3)/4},\quad &\hat\nu_c<\nu
        \end{aligned}
        \right.\,,
    \end{eqnarray}
for slow cooling. The broken power-law/broken power-law division in Equation~\ref{eq:N_ic} and \ref{eq:syn_ic} give rise to additional spectral segments above $\gamma_c$ ($\nu_c$). As shown in Figure~\ref{fig:spectrum}, the blue lines correspond to synchrotron-only cooling, the green lines to Thomson IC cooling with constant $\tilde{Y}$, and the orange lines to spectra including Klein–Nishina corrections, which introduce additional breaks for $\nu > \nu_c$.

\subsection{Inverse Compton Spectrum}   

\begin{figure*}
        \centering
        \includegraphics[width=0.33\textwidth]{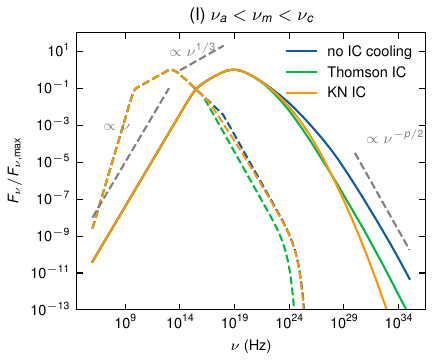}
        \includegraphics[width=0.33\textwidth]{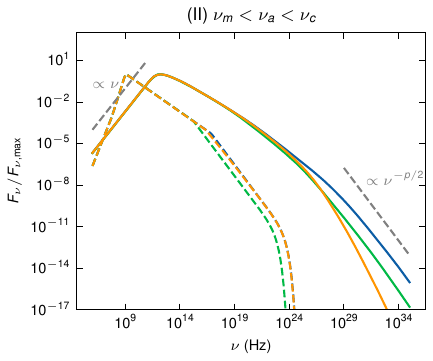}
        \includegraphics[width=0.33\textwidth]{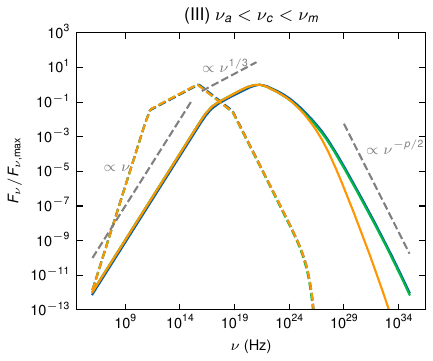}\\
        \includegraphics[width=0.33\textwidth]{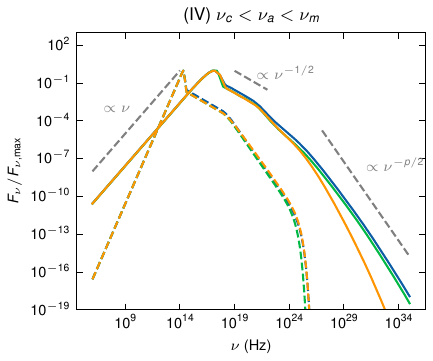}
        \includegraphics[width=0.33\textwidth]{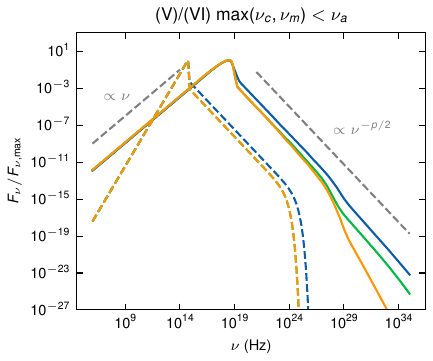}
        \caption{Five characteristic spectral shapes of observed self-synchrotron Compton (SSC) emission from an on-axis narrow top-hat jet are shown. Blue curves represent spectra without inverse Compton cooling. Green curves include inverse Compton cooling (omitting Klein–Nishina corrections), while orange curves show the effect of including full Klein–Nishina corrections in the inverse Compton cooling.}
        \label{fig:IC-spectrum}
\end{figure*}

We calculate the spectrum of the upscattered photon (IC spectrum) numerically from \citep{Rybicki1979,Sari2001},
\begin{eqnarray}
    I_\nu^{\rm IC}(\nu)&=&\delta^\prime\int_{1}^\infty\frac{dn_e}{d\gamma}d\gamma\int_0^{x_0}\sigma_c(x)I_{\nu_0}^{\rm syn}(x)dx\\
      &=&\int_{1}^\infty\frac{d\Sigma_e}{d\gamma}d\gamma\int_0^{\frac{\nu}{4\gamma^2x_0}}\sigma_c(\frac{\gamma h\nu_0}{m_ec^2})I_{\nu_0}^{\rm syn}(\nu_0)\frac{\nu d\nu_0}{4\gamma^2\nu_0^2}\nonumber\,,
    \end{eqnarray}
    where 
\begin{eqnarray}
        \sigma_c &=& \frac{3}{4}\sigma_T\bigg[\frac{1+x}{x^3}\bigg\{\frac{2x(1+x)}{1+2x}-\ln(1+2x)\bigg\}\,,\\
        &+&\frac{1}{2x}\ln(1+2x)-\frac{1+3x}{(1+2x)^2} \bigg]\\
        x&=&\frac{h\nu}{m_ec^2}\,,
\end{eqnarray}
and $\delta^\prime$ is the co-moving shell width. This method is valid for an arbitrary input electron spectrum and photon spectrum and includes Klein-Nishina corrections in the higher-energy photon regime. Some approximate analytical expressions of the self-synchrotron Compton (SSC) spectrum without the Klein-Nishina correction have been derived from synchrotron cooling only electron/photon spectra (Section~\ref{sec:e_spectrum} and Section~\ref{sec:syn_rad}) in six different regimes \citep{gao_13}. However, these approximations exhibit unphysical spectral discontinuities and light curve jumps near the transition frequencies.

Figure~\ref{fig:IC-spectrum} shows the calculated SSC spectrum and corresponding synchrotron spectrum in six different regimes with different cooling treatments. The blue lines show the results from synchrotron cooling only, while the green lines include IC cooling but only for Thompson cooling and cross-section. The orange lines show the complete results that include both KN-corrected IC cooling and Compton cross-section, where the high-energy portion of the spectrum gets suppressed. The gray dashed lines are power-law scalings derived in \citep{gao_13}. Scalings in the intermediate regions are omitted as the SSC spectrum no longer follows a simple power-law in this range.

\section{Geometry}\label{sec:geometry}
The radiations of each fluid grid are first calculated in the fluid co-moving frame as described in Section~\ref{sec:radiation}. For an observer with given viewing angles $\theta_v$ and $\phi_v$, Doppler transformation of the radiation in each grid cell from the co-moving frame to the observer frame is performed. The Doppler factor is written as
\begin{eqnarray}
    \mathcal{D}=\frac{1}{\Gamma(1-\beta\cos w)}\,,
\end{eqnarray}
where
\begin{eqnarray}
    \cos w= \sin\theta\cos(\phi-\phi_v)\sin\theta_v+\cos\theta\cos\theta_v\,,
\end{eqnarray}
is the cosine of the angular distance between the line of sight and the target cell. Then the observed specific flux for a given time can be calculated by adding all cells along the equal arrival time surface,

\begin{eqnarray}
        F_\nu(\nu,\tilde{t})=\frac{1+z}{4\pi D_L^2}\oiint_{t_{\rm obs}=\tilde{t}}  4\pi r^2 I_{\nu^\prime}(\nu^\prime)\mathcal{D}^3d\Omega dr\,,
\end{eqnarray}
where
\begin{eqnarray}
    \nu^\prime &=& \frac{1+z}{\mathcal{D}}\nu\,,\\
    t_{\rm obs}&=&(1+z)\bigg[t+\frac{r}{c}(1-\cos w)\bigg]\,.
\end{eqnarray}

\section{Examples}\label{sec:eg}

\begin{table*}[htbp]
\centering
\footnotesize
\caption{Feature comparison of some public afterglow modeling codes.}
\begin{tabular}{@{}llccccc@{}}
\toprule
\textbf{Category} & \textbf{Feature} & \href{https://github.com/geoffryan/afterglowpy}{\textbf{Afterglowpy}} & \href{https://github.com/haowang-astro/jetsimpy}{\textbf{JetSimPy}} & \href{https://github.com/leiwh/PyFRS}{\textbf{PyFRS}} & \href{https://github.com/vsevolodnedora/PyBlastAfterglowMag}{\textbf{PyBlastAfterglow}} & \href{https://github.com/YihanWangAstro/VegasAfterglow}{\textbf{VegasAfterglow}} \\
 & & \cite{ryan_20} & \cite{jetsimpy2025} & \cite{Lei2016} & \cite{Nedora2024} & This paper \\
\midrule
\textbf{General} 
    & Backend language         & C          & C++        & Python     & C++        & C++ \\
    & Interface language       & Python     & Python     & Python     & Python     & Python/C++ \\

\midrule
\textbf{Jets} 
    & Structured jet           & \cmark     & \cmark     & \xmark     & \cmark     & \cmark \\
    & Jet magnetization        & \xmark     & \xmark     & \xmark     & \xmark     & \cmark \\
    & Lateral spreading        & \cmark     & \cmark     & \xmark     & \cmark     & \cmark \\
    & Non-axisymmetric jet     & \xmark     & \xmark     & \xmark     & \xmark     & \cmark \\
    & Arbitrary viewing angle  & \cmark     & \cmark     & \cmark     & \cmark     & \cmark \\

\midrule
\textbf{Dynamics} 
    & Reverse shock            & \xmark     & \cmark     & thin shell & thin shell & \cmark \\
    & Deep Newtonian correction& optional   & optional   & \cmark     & \cmark     & \cmark \\
    & Radiative fireball       & \xmark     & \xmark     & \cmark     & \cmark     & \cmark \\
    & Medium type              & ISM        & Arbitrary   & ISM/Wind   & ISM/Wind   & Arbitrary \\
    & Energy injection         & \cmark     & \cmark     & \cmark     & \xmark     & \cmark \\
    & Matter injection         & \xmark     & \cmark     & \xmark     & \xmark     & \cmark \\

\midrule
\textbf{Radiation} 
    & Self-absorption          & \xmark     & \xmark     & optical depth  & optical depth   & black body \\
    & IC cooling               & \cmark     & \xmark     & \xmark     & \cmark     & \cmark \\
    & IC radiation          & \xmark     & \xmark     & \cmark     & \cmark     & \cmark \\
    & Klein-Nishina correction & \xmark     & \xmark     & \xmark     & \cmark     & \cmark \\
\bottomrule
\label{tab:code}
\end{tabular}
\end{table*}
In this section, we show several multi-wavelength light-curve examples generated by {\tt VegasAfterglow} and compare them with the results from other public codes implementing the same features. Table~\ref{tab:code} summarizes the features implemented in various public codes to date. Most codes offer jet–structure modeling, whereas treatments of reverse‐shock dynamics (especially in the thick‐shell regime) and inverse‐Compton emission with Klein–Nishina corrections are comparatively underdeveloped.

\subsection{Strctured Jets}
\begin{figure*}
        \centering
        \includegraphics[width=1\textwidth]{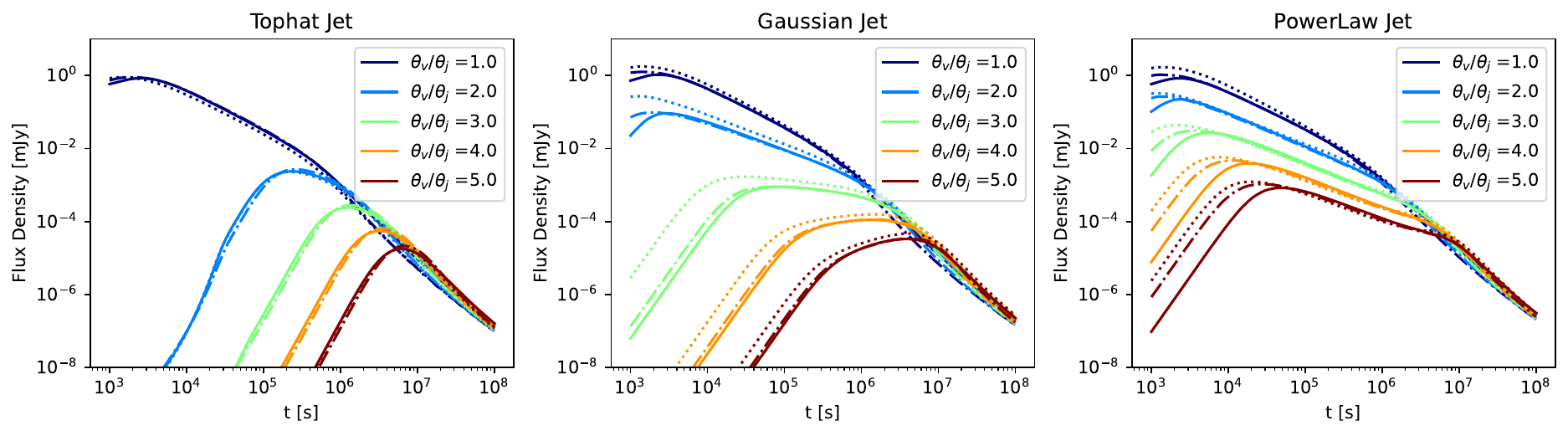}
        \caption{Effect of viewing angle geometry on optical light curves of different viewing angles for non-spreading top-hat, Gaussian, and power-law jet structures with $E_{\rm k, iso}=10^{53}$ erg, $\Gamma_0=300$, $\theta_j=5^\circ$, $n_{\rm ism}=1$ cm$^{-3}$, $p=2.3$, $\epsilon_e=10^{-1}$, $\epsilon_B=10^{-3}$ and $z=1$. Solid lines: {\tt VegasAfterglow}; Dash dotted lines: {\tt Jetsimpy}; Dotted lines: {\tt AfterglowPy}.}
        \label{fig:viewing-angle}
\end{figure*}
First, we present examples of structured jets observed from various viewing angles and compare them with the results from {\tt AfterglowPy} and {\tt Jetsimpy} \citep{wang_etal_24}. The left panel of Figure~\ref{fig:viewing-angle} shows the light curve for a top-hat jet with a half-opening angle of $5^\circ$. All three codes agree closely in this simple case; however, {\tt AfterglowPy} exhibits a slight deviation at small viewing angles during the early coasting phase due to its approximate treatment of that regime. The middle panel illustrates a Gaussian jet with a $5^\circ$ core angle, while the right panel shows a power-law jet with index 2 and the same core angle. For structured jets, both {\tt AfterglowPy} and {\tt Jetsimpy} predict higher fluxes. This arises from their different treatments of the angular Lorentz‐factor profile: {\tt AfterglowPy} allows $\Gamma(\theta)$ to fall below unity but truncates emission beyond a “wing” angle $\theta_{\rm wing}$, whereas {\tt Jetsimpy} and {\tt VegasAfterglow} model $\Gamma - 1$ with a Gaussian profile and apply the “+1” correction.

\subsection{Deep Newtonian Regime}
\begin{figure}
        \centering
        \includegraphics[width=0.45\textwidth]{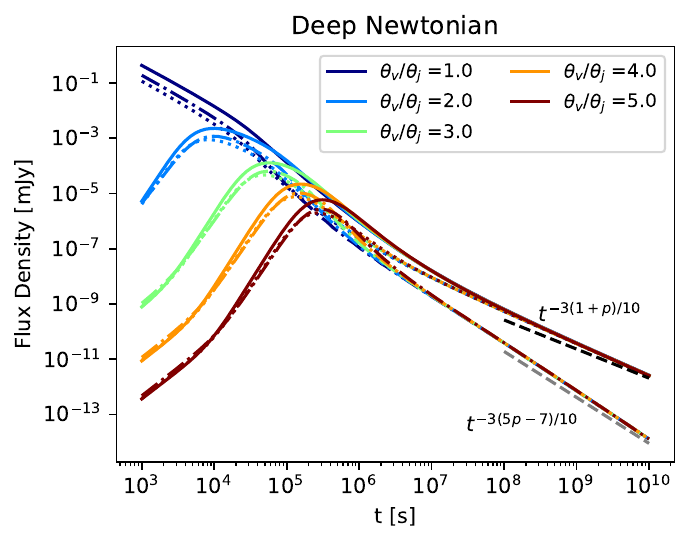}
        \caption{Effect of deep Newtonian on optical light curves of different viewing angles for a non-spreading top-hat with $E_{\rm k, iso}=10^{53}$ erg, $\Gamma_0=300$, $\theta_j=5^\circ$, $n_{\rm ism}=10^4$ cm$^{-3}$, $p=2.5$, $\epsilon_e=10^{-3}$, $\epsilon_B=10^{-3}$ and $z=1$. Solid lines: {\tt VegasAfterglow}; Dash dotted lines: {\tt Jetsimpy}; Dotted lines: {\tt AfterglowPy}.}
        \label{fig:deep-newtonian}
\end{figure}
Figure~\ref{fig:deep-newtonian} shows optical ($10^{14}$ Hz) light curves for a tophat jet propagating in a very dense ISM with $n_{\rm ism}=10^4$ cm$^{-3}$ for various viewing angles in the late‐time, deep‐Newtonian regime. {\tt VegasAfterglow} applies the deep‐Newtonian correction by default, whereas {\tt AfterglowPy} and {\tt Jetsimpy} offer it as an optional feature.  The plot demonstrates that the deep‐Newtonian correction flattens the light curve at late times. {\tt VegasAfterglow} and {\tt AfterglowPy} predict $\propto t^{-3(1+p)/10}$ in the deep Newtonian regime based on the assumption that the momentum of the shocked electron follows the power-law distribution discussed in Section~\ref{sec:gamma_m}, while {\tt Jetsimpy} predicts $\propto t^{-3(5p-7)/10}$ based on the assumption that the energy of the shocked electron follows the power-law distribution.  Note also that {\tt VegasAfterglow} predicts systematically higher fluxes than the other two codes; this arises from the “+1” correction described in Section~\ref{sec:gamma_m}, which increases the characteristic frequency $\nu_m$.  If the observed frequency is above $\nu_m$, which is the case here, the observed flux will be higher. This effect becomes significant at low bulk Lorentz factors $\Gamma$, considered here for the late stage.

\subsection{Synchrotron Self-absorption}
\begin{figure}
        \centering
        \includegraphics[width=0.45\textwidth]{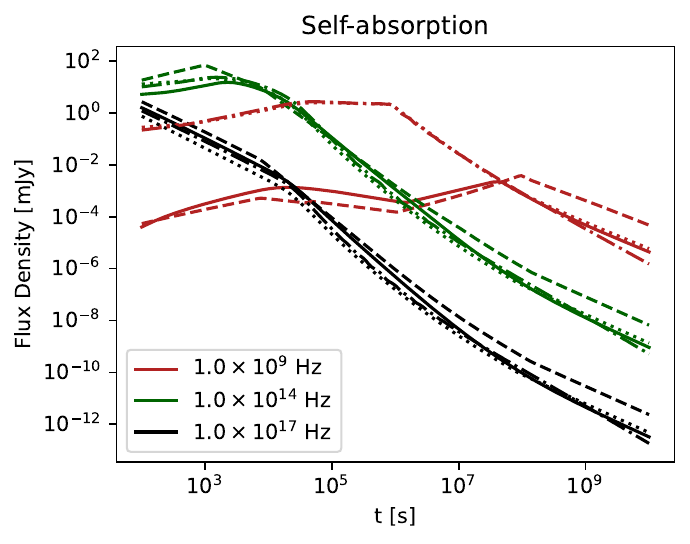}
        \caption{Multi-wavelength light curves for a non-spreading top-hat with $E_{\rm k, iso}=10^{53}$ erg, $\Gamma_0=300$, $\theta_j=5^\circ$, $n_{\rm ism}=10^2$ cm$^{-3}$, $p=2.3$, $\epsilon_e=10^{-1}$, $\epsilon_B=10^{-3}$ and $z=1$. Solid lines: {\tt VegasAfterglow}; Dash dotted lines: {\tt Jetsimpy}; Dotted lines: {\tt AfterglowPy}; Dashed lines: {\tt PyFRS}.}
        \label{fig:self-absorption}
\end{figure}
Figure~\ref{fig:self-absorption} compares synchrotron self-absorption in {\tt VegasAfterglow} with three other codes: {\tt Jetsimpy}, {\tt AfterglowPy}, and {\tt PyFRS}.  Neither {\tt Jetsimpy} nor {\tt AfterglowPy} include self-absorption, whereas {\tt PyFRS} treats it via an optical depth method. As shown in the figure, results from the four codes agree well in the optical and X-ray bands, but in the radio band both {\tt VegasAfterglow} and {\tt PyFRS} (with self-absorption) predict substantially lower fluxes than the other two codes that omit this effect.

\subsection{Radiative Fireball}
\begin{figure}
        \centering
        \includegraphics[width=0.45\textwidth]{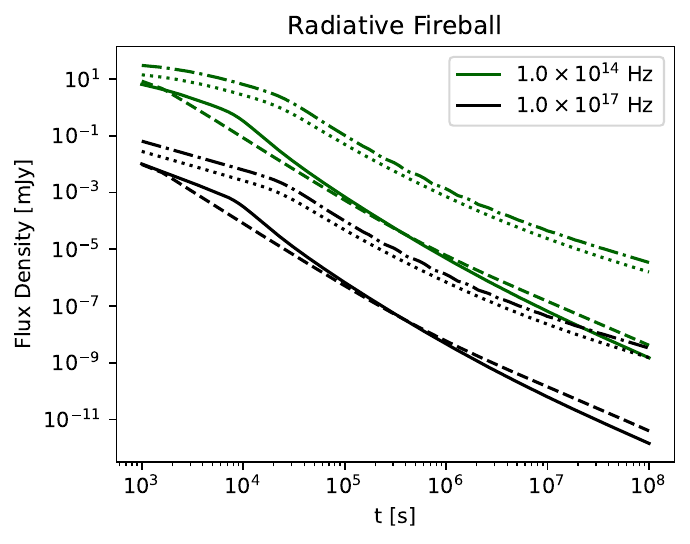}
        \caption{Multi-wavelength light curves for a non-spreading top-hat with $E_{\rm k, iso}=10^{53}$ erg, $\Gamma_0=300$, $\theta_j=5^\circ$, $n_{\rm ism}=10^2$ cm$^{-3}$, $p=2.001$, $\epsilon_e=0.9$, $\epsilon_B=0.05$ and $z=1$. Solid lines: {\tt VegasAfterglow}; Dash dotted lines: {\tt Jetsimpy}; Dotted lines: {\tt AfterglowPy}; Dashed lines: {\tt PyFRS}.}
        \label{fig:radiative}
\end{figure}
As discussed in Section~\ref{sec:forward}, when the electron energy fraction $\epsilon_e$ is high (e.g., $\sim 1$) and the blast wave enters the fast-cooling regime ($\nu_c < \nu_m$), the system becomes radiative. In this case, the shock dynamics deviate significantly from the adiabatic scenario (where $\epsilon_{\rm rad} = 0$), which is commonly assumed in most afterglow models. Figure~\ref{fig:radiative} presents the optical and X-ray light curves for a radiative blast wave. Compared to the adiabatic case, the emission is markedly fainter and exhibits an earlier jet break, primarily due to the more rapid decrease in the Lorentz factor caused by radiative energy losses.

\subsection{Reverse Shock Emission}
\begin{figure*}
        \centering
        \includegraphics[width=0.95\textwidth]{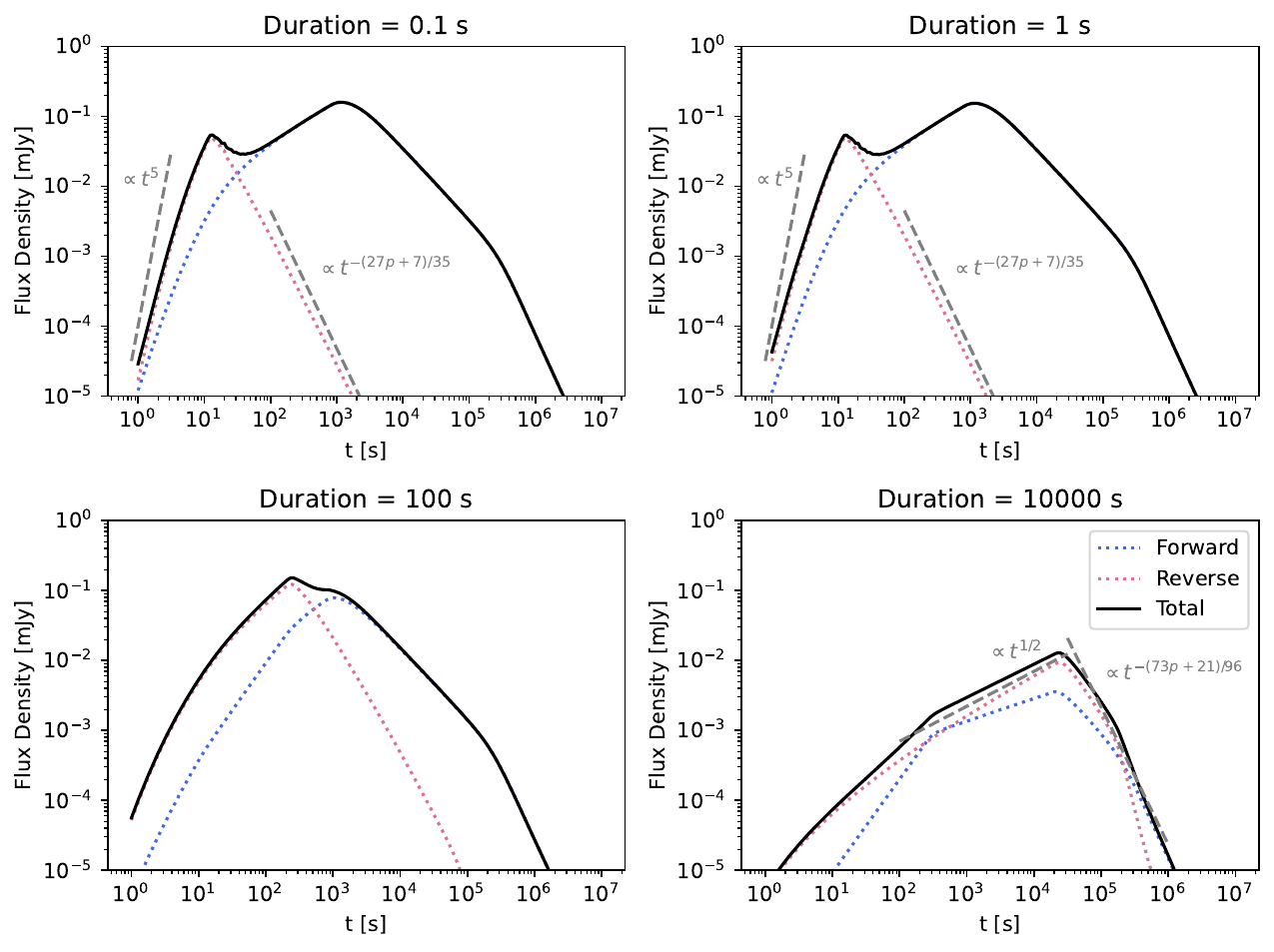}
        \caption{Optical forward and reverse shock light curves for a non-spreading top-hat with $E_{\rm k, iso}=10^{52}$ erg, $\Gamma_0=300$, $\theta_j=10^\circ$, $n_{\rm ism}=1$ cm$^{-3}$, $p_f=p_r=2.3$, $\epsilon_{e,r}=\epsilon_{e,f}=0.1$, $\epsilon_{B,r}=\epsilon_{B,f}=10^{-3}$ and $z=1$. }
        \label{fig:reverse}
\end{figure*}
Figure~\ref{fig:reverse} presents examples of on-axis optical ($10^{14}$ Hz) light curves from the forward and reverse shocks under different top-hat jet injection durations (with identical other parameters). The blue dashed lines represent the forward shock emission, while the red dashed lines correspond to the reverse shock. When the injection duration $\tau$ is much shorter than the deceleration time (e.g., $0.1$ s and $1$ s in this case), the reverse shock operates in the thin-shell regime. In this regime, the reverse shock crosses the entire ejecta shell around the deceleration time and reaches the same peak $\Gamma_{43}\sim1.1$, indicated in the upper right panel of Figure~\ref{fig:r_dynamics}. This results in nearly identical reverse shock light curves, as seen in the upper panels. The reverse shock emission scaling deviates from its $\propto t^5$ dependence near the shock crossing time as the system transitions out of the deep Newtonian regime. This transition causes the Lorentz factors $\Gamma_3$ and $\Gamma_4$ to diverge. As a result, the scaling of the relative Lorentz factor, $\Gamma_{43}$-1, also fails to follow the derived $\propto t^3$ relation, as demonstrated in the upper right panel of Figure~\ref{fig:r_dynamics}.

As the injection duration increases and becomes comparable to the deceleration time, the system transitions to the thick-shell regime. The reverse shock crossing takes significantly longer, leading to a delayed peak in the reverse shock emission. Additionally, the prolonged reverse shock interaction further decelerates the jet shell, reducing the Lorentz factor during the reverse shock crossing phase (See first panel of Figure~\ref{fig:r_dynamics}) and resulting in fainter forward shock emission, as shown in the lower panels.
In the thick shell regime, the post-crossing reverse shock emission scaling $\propto t^{-(73p+21)/96}$ is derived based on the assumption $g=7/2$. However, in \texttt{VegasAfterglow}, we adopt a ``temperature"-dependent $g$, as shown in Figure~\ref{fig:g}. Consequently, a small deviation from this post-crossing scaling is observed in the bottom right panel. The bottom left panel, being in the transition regime between the thin-shell and thick-shell regimes, does not allow for the application of any analytical scaling.

\subsection{Energy Injection}
\begin{figure}
        \centering
        \includegraphics[width=0.45\textwidth]{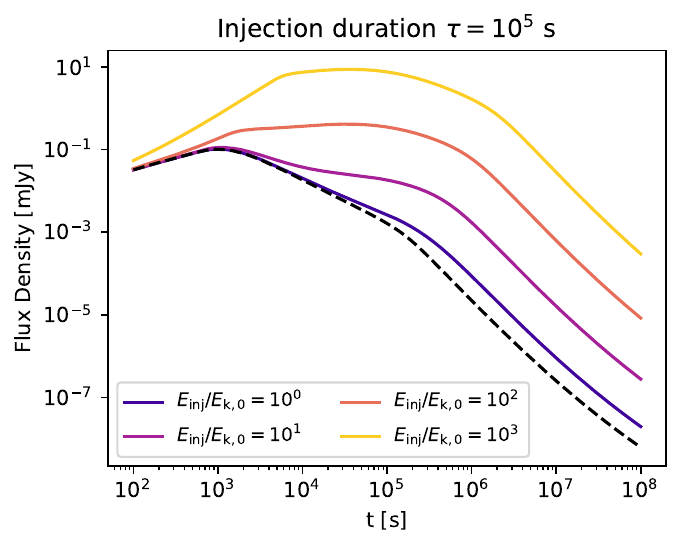}
        \caption{Example of energy injection into an on-axis top-hat jet. Optical ($10^{14}$ Hz) light curves are shown for different total injection energies, with the injection timescale $\tau$ fixed at 10$^5$ s for visual comparison. $E_{\rm k, iso}=10^{52}$ erg, $\Gamma_0=300$, $\theta_j=10^\circ$, $n_{\rm ism}=1$ cm$^{-3}$, $p=2.3$, $\epsilon_e=10^{-1}$, $\epsilon_B=10^{-3}$ and $z=1$.}
        \label{fig:energy-inj}
\end{figure}
Figure~\ref{fig:energy-inj} shows an example of on-axis optical light curves of energy injection to a tophat jet with different total injection energies, mimicking the same injection function as magnetar spin-down as described in Equation~\ref{eq:magnetar}. We fix the injection duration to be $10^5$ s for better visual comparison. A plateau feature is formed as the total injected energy is much larger than the initial kinetic energy of the jet.

\subsection{User-defined jet}
\begin{figure}
        \centering
        \includegraphics[width=0.5\textwidth]{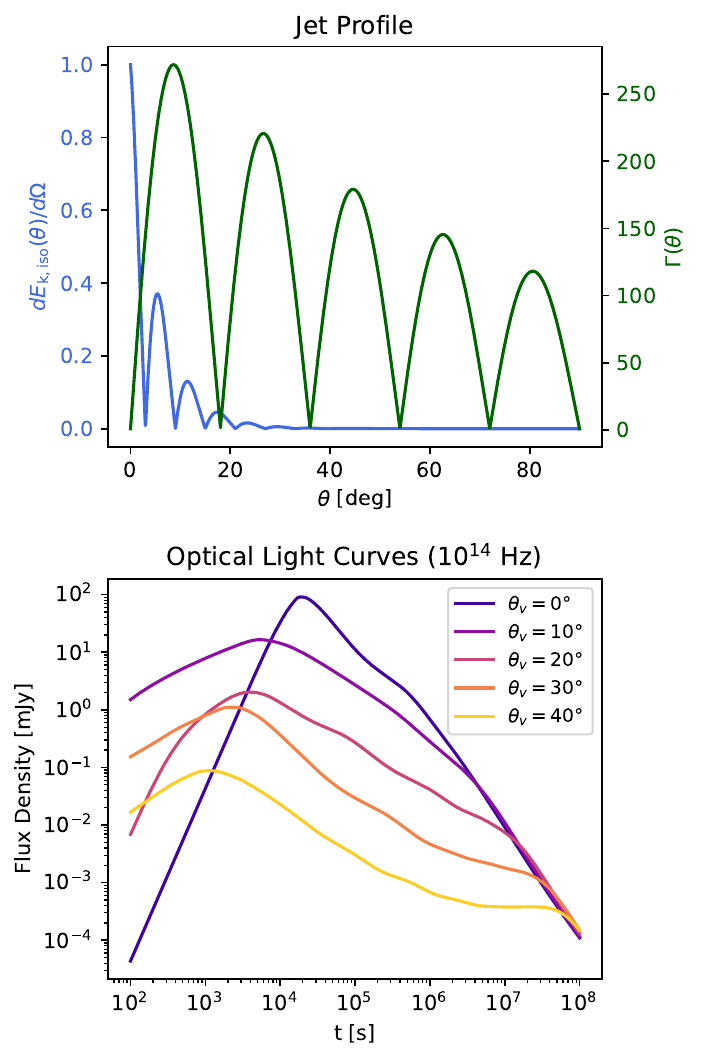}
        \caption{An example user-defined jet. The upper panel shows the energy and Lorentz factor profile while the bottom panel shows the optical ($10^{14}$ Hz) light curves from different viewing angles.}
        \label{fig:user}
\end{figure}
Figure~\ref{fig:user} illustrates a random example of a user-defined jet where the energy and Lorentz factor profiles are specified by Python functions. The upper panel presents an example using a damped cosine function for the energy profile and a damped sine function for the Lorentz factor profile. The bottom panel displays the corresponding optical ($10^{14}$ Hz) light curves calculated for several viewing angles. The {\tt VegasAfterglow} code supports any arbitrary functions of $\phi$ and $\theta$ for energy, Lorentz factor, magnetization, and arbitrary functions of $\phi$, $\theta$, and $t$ for energy injection and matter injection, as detailed in Section~\ref{sec:jet}.

\subsection{Performance Tests}
\begin{figure}
        \centering
        \includegraphics[width=0.45\textwidth]{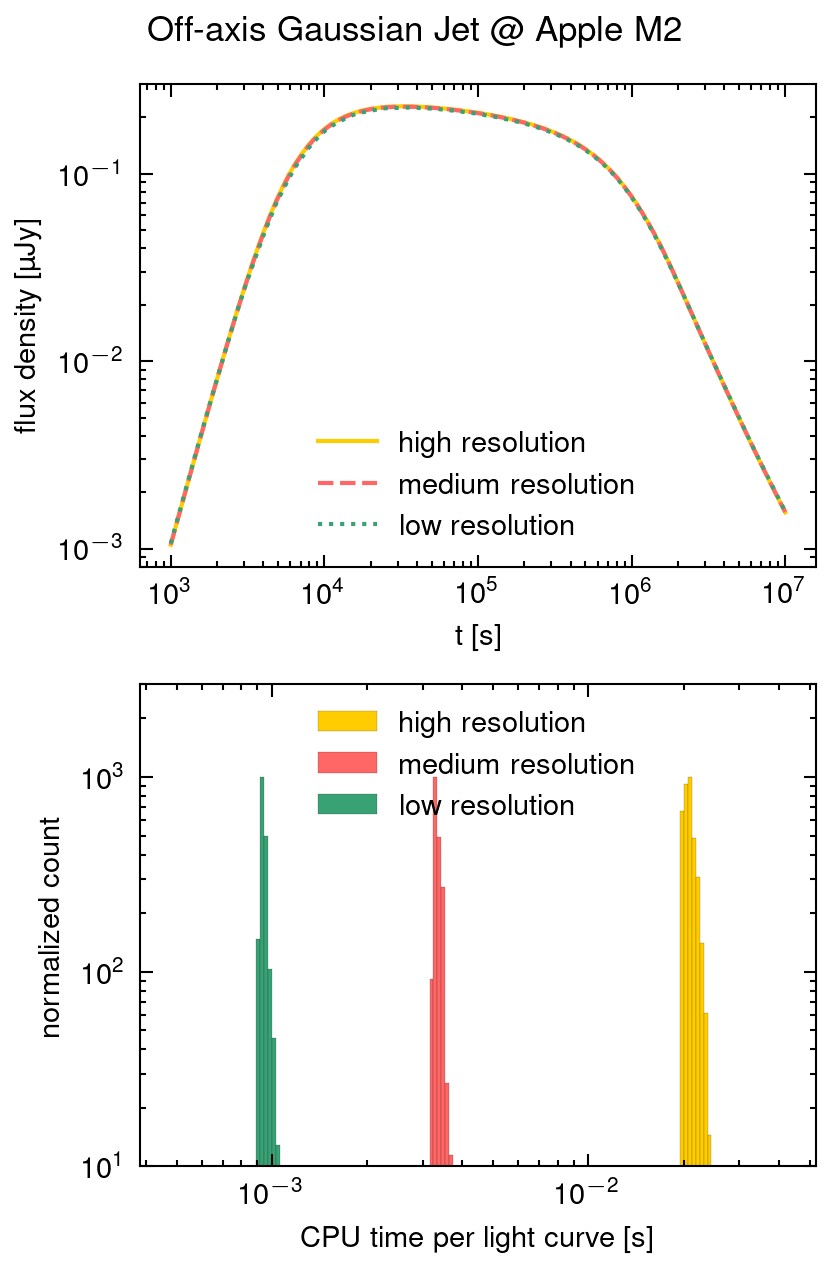}
        \caption{Convergence and performance test of {\tt VegasAfterglow} on an Apple M2 laptop. The upper panel shows the optical light curve from an off-axis Gaussian jet computed with varying grid resolutions. The lower panel presents the average CPU time per light curve for 10,000 realizations with parameters slightly perturbed from those used in the upper panel. The fiducial parameters are: $E_{\rm k, iso} = 10^{52}$ erg, $\Gamma_0 = 300$, $\theta_j = 0.1$ rad, $\theta_v = 0.3$ rad, $n_{\rm ism} = 1$ cm$^{-3}$, $p = 2.3$, $\epsilon_e = 10^{-2}$, $\epsilon_B = 10^{-4}$, and $z = 0.009$.}
        \label{fig:performance}
\end{figure}
{\tt VegasAfterglow} is highly optimized at the algorithmic level to achieve high performance in light curve generation. Figure~\ref{fig:performance} presents a convergence and performance test using an off-axis Gaussian jet. The fiducial parameters are: $E_{\rm k, iso} = 10^{52}$ erg, $\Gamma_0 = 300$, $\theta_j = 0.1$ rad, $\theta_v = 0.3$ rad, $n_{\rm ism} = 1$ cm$^{-3}$, $p = 2.3$, $\epsilon_e = 10^{-2}$, $\epsilon_B = 10^{-4}$, and $z = 0.009$. The upper panel shows the convergence behavior with varying grid resolutions, while the lower panel displays the average CPU time per light curve for $10,000$ realizations with parameters slightly perturbed from the fiducial setup.

For this fiducial configuration and forward-shock synchrotron-only calculation, {\tt VegasAfterglow} can generate a single-frequency light curve in approximately $1$ ms when using sufficiently low resolution while maintaining convergence with higher resolutions. The benchmark is performed at the C++ level and can be reproduced using the script provided in our public code repository. Although the grid resolution required to achieve numerical convergence may be problem-dependent, {\tt VegasAfterglow} nevertheless remains extremely fast for typical GRB afterglow modeling scenarios.

This high level of performance is the result of targeted C++ optimizations in algorithmic design, data structures, and memory access patterns. While a comprehensive discussion of software engineering is beyond the scope of this paper, we highlight several key optimizations below.

\begin{itemize}
    \item log-log interpolations on grids: Due to the power-law nature of afterglow physics, many quantities exhibit linear behavior in log-log space. By performing interpolation in logarithmic space, {\tt VegasAfterglow} achieves significantly higher accuracy with fewer grid points compared to linear interpolation in linear space.
    \item Avoiding redundant power-law calculations: Spectral calculations often involve repeated power-law evaluations. The built-in power function is computationally expensive, typically requiring tens to hundreds of CPU cycles per call---much more than basic arithmetic operations. By leveraging log-log interpolation, power-law evaluations can be replaced with simple additions, subtractions, and multiplications, greatly accelerating light curve computations.
    \item Sequential memory access patterns: The code is designed to favor sequential memory access, minimizing cache misses and improving data locality. This ensures that critical operations can take full advantage of modern CPU caching mechanisms, leading to better memory throughput and overall runtime performance.
\end{itemize}

\section{Summary and conclusions}
\label{sec:conclusion}
We have presented \texttt{VegasAfterglow}, a high-performance C++ framework with Python interfaces for flexible and accurate modeling of GRB afterglows. The framework combines advanced numerical methods with efficient implementation, enabling fast and reliable modeling across a broad range of physical scenarios. Below, we summarize the key features implemented in the current version of the code, followed by the main scientific insights derived from our studies.
\subsection{Framework Capabilities}
\begin{itemize}
 \item \textbf{Comprehensive shock dynamics:} \texttt{VegasAfterglow} self-consistently solves the dynamics of both forward and reverse shocks, including support for arbitrary upstream magnetization, structured jets, and user-defined ambient density profiles. The implementation allows for a seamless transition between the thin-shell and thick-shell regimes, ensuring accurate modeling across the full range of ejecta conditions.

\item \textbf{General jet structure support:} Users can define custom profiles for jet energy, Lorentz factor, and magnetization, as well as energy and mass injection histories, allowing flexible modeling of a wide range of jet geometries, including non-axisymmetric configurations.

\item \textbf{Full radiation treatment:} The framework calculates both synchrotron and inverse Compton radiation, including synchrotron self-absorption in all spectral regimes and Klein–Nishina corrections for inverse Compton, and supports arbitrary observing angles and frequency grids.

\item \textbf{Full velocity regimes:} The code operates seamlessly across relativistic and trans-relativistic dynamical regimes, incorporating lateral jet spreading and transitions to deep-Newtonian regime.

\item \textbf{Radiative Fireball:} We adopt a more accurate treatment of shock dynamics that incorporates both adiabatic expansion and radiative energy losses. This allows for a smooth transition between the adiabatic and radiative fireball regimes, a feature often neglected in other publicly available GRB afterglow modeling codes. As a result, dimmer afterglow light curves are expected in cases with large electron energy fractions ($\epsilon_e$).
  
\item \textbf{High computational efficiency:} The implementation is highly optimized to deliver ultra-fast light curve and spectrum calculations, enabling comprehensive Bayesian inference on standard laptop hardware within seconds to minutes, rather than the hours or days typically required. On a 2022 Apple M2 chip, a single-frequency light curve with 100 time points can be computed in approximately 1 ms for synchrotron emission from a structured jet viewed off-axis, using a modest grid resolution sufficient for convergence. This accelerated performance facilitates rapid iteration over physical models, making the framework well-suited for both detailed analyses of individual GRB events and large-scale population studies.

\item \textbf{Public availability and modular design:} The framework is open-source and publicly available at \url{https://github.com/YihanWangAstro/VegasAfterglow}, designed with modular components to facilitate extension and integration into broader workflows.
\end{itemize}

\subsection{Scientific Findings}
\begin{itemize}
  \item We derive, for the first time, a general analytical shock jump condition valid for arbitrary upstream magnetization $\sigma_{\rm u}$ and relative Lorentz factor $\Gamma_{\rm ud}$. The key result, providing the downstream four-velocity, is presented in Equation~\ref{eq:u_down}. In the limit $\sigma_{\rm u} \to 0$, this solution reduces to the classical Blandford-McKee result, while in the high-$\sigma$, highly relativistic regime, it asymptotically approaches the Kennel-Coroniti solution. This general formulation enables consistent modeling of both forward and reverse shocks across the full range of magnetization and Lorentz factor regimes.

  \item We introduce a unified treatment of reverse shock crossing dynamics that self-consistently handles all shell thicknesses, enabling a seamless transition between the thin-shell and thick-shell regimes for arbitrary magnetizations. This new approach also preserves the energy conservation of the blast wave, similar to the so-called mechanical model, but with significantly reduced complexity. Instead of performing detailed integrations over the blast wave profile, our method relies on particle number conservation, providing a simpler yet physically consistent alternative.

  \item Our simulations of detailed reverse shock crossing dynamics reveal that, in the thin-shell regime, reverse shocks are systematically weaker than previously estimated by analytical approximation in the literature. Specifically, we find that the peak relative Lorentz factor is typically { $\Gamma_{43} \sim 1.02$}, significantly lower than the commonly cited value of $7/4$. This suggests that predictions of reverse shock emission based on earlier analytical scalings, which rely on $\Gamma_{43}-1$, are likely overestimated.

\end{itemize}

\section*{Acknowledgements}
We thank all contributors who helped improve \texttt{VegasAfterglow} during its development. In particular, we are grateful to Weihua Lei, Shaoyu Fu, Iris Yin, Cuiyuan Dai, Liang-Jun Chen and Binbin Zhang for their invaluable efforts as beta testers, providing constructive feedback and assisting with bug fixes. The authors acknowledge NASA 80NSSC23M0104, NASA 80NSSC20M0043, the Nevada Center for Astrophysics for support.

    \bibliographystyle{elsarticle-harv}
    \bibliography{main}

\end{document}